\documentclass[12pt,a4paper]{article}
\pdfoutput=1
\usepackage{color}
\usepackage{amssymb,amsmath,bm,bbm}
\usepackage{epsf}
\usepackage{epsfig}
\usepackage{afterpage}
\usepackage{longtable}
\usepackage{booktabs}
\usepackage{caption}
\usepackage[dvipsnames]{xcolor}
\usepackage[linktoc=page,bookmarks=false,colorlinks=false,
linkbordercolor=RoyalBlue,citebordercolor=ForestGreen,urlbordercolor=CornflowerBlue]{hyperref}
\usepackage{latexsym,mathrsfs,dsfont}
\usepackage[normalem]{ulem} % for strikeout \sout
\usepackage[compress]{cite}
\usepackage{graphicx}
\usepackage{url}
\usepackage{paralist}
\usepackage{bbold}

\newcommand{\f}{\frac}
\newcommand{\ord}{\mathcal{O}}
\def\as{\alpha_s}
\def\aspi{\frac{\as}{4\pi}}

\newcommand{\gev}{\, {\rm GeV}}
\newcommand{\mev}{\, {\rm MeV}}

\newcommand{\bsi}{B_6^{(1/2)}}
\newcommand{\bei}{B_8^{(3/2)}}

\def\epe{\varepsilon'/\varepsilon}
\newcommand{\beq}{\begin{equation}}
\newcommand{\eeq}{\end{equation}}
\newcommand{\be}{\begin{equation}}
\newcommand{\ee}{\end{equation}}
\newcommand{\bi}{\begin{itemize}}
\newcommand{\ei}{\end{itemize}}
\newcommand{\ba}{\begin{array}}
\newcommand{\ea}{\end{array}}
\newcommand{\beqa}{\begin{eqnarray}}
\newcommand{\eeqa}{\end{eqnarray}}
\newcommand{\bea}{\begin{eqnarray}}
\newcommand{\eea}{\end{eqnarray}}
\newcommand{\beqn}{\begin{eqnarray}}
\newcommand{\eeqn}{\end{eqnarray}}

\definecolor{red}{cmyk}{0,1,1,0.4}

\usepackage{fancyhdr}
\advance \headheight by 15.0truept       % for 12pt mandatory...
\begin{document}

%\begin{flushleft}
%{\em Version of 20 July 2015}
%\end{flushleft}

\vspace{-14mm}
\begin{flushright}
        {AJB-18-5}\\
CP3-18-31
\end{flushright}

\vspace{6mm}

\begin{center}
{\Large\bf
\boldmath{BSM Hadronic Matrix Elements for $\epe$ and $K\to\pi\pi$ Decays in the Dual QCD Approach}}
\\[12mm]
{\bf \large Jason Aebischer${}^a$, Andrzej~J.~Buras${}^b$ and Jean-Marc G\'erard${}^c$ \\[0.8cm]}
{\small
${}^a$Excellence Cluster Universe, TUM, Boltzmannstr.~2, 85748~Garching, Germany \\[2mm]
jason.aebischer@tum.de \\[2mm]
${}^b$TUM Institute for Advanced Study, Lichtenbergstr.~2a, D-85748 Garching, Germany\\
Physik Department, TU M\"unchen, James-Franck-Stra{\ss}e, D-85748 Garching, Germany\\[2mm]
aburas@ph.tum.de\\[2mm]
${}^c$ Centre for Cosmology,
Particle Physics and Phenomenology (CP3), Universit{\'e} catholique de Louvain,
Chemin du Cyclotron 2,
B-1348 Louvain-la-Neuve, Belgium}\\[2mm]
jean-marc.gerard@uclouvain.be
\end{center}
\begin{center}
 \today
\end{center}

\vspace{6mm}

\abstract{%
\noindent
We calculate for the first time all four-quark  hadronic matrix elements  of local operators possibly contributing to $K\to\pi\pi$ decays and in particular to the ratio $\epe$ beyond the Standard Model (BSM). To this end we use the Dual QCD (DQCD) approach. In addition to 7 new  mirror operators obtained from the SM ones by flipping the chirality, we count 13 BSM four-quark operators of a given chirality linearly independent of each other and of the aforesaid 14 operators for which hadronic matrix elements are already known. We present results in two bases for all these operators, one
termed DQCD basis useful for the calculation of the hadronic matrix elements in
 the DQCD approach and the other called SD basis suited to the short distance renormalization group  evolution above the $1\gev$ scale.
We demonstrate that the pattern of long
distance evolution (meson evolution) matches the one of short distance evolution (quark-gluon evolution), a property which to our knowledge cannot be presently achieved in any other analytical framework.  The highlights of our paper
are chirally enhanced matrix elements of tensor-tensor and scalar-scalar BSM
operators.
They could thereby explain the emerging $\epe$ anomaly which is strongly indicated within DQCD with some support from lattice QCD. On the other hand
we do not expect the BSM operators to be relevant for the $\Delta I=1/2$ rule.}

\setcounter{page}{0}
\thispagestyle{empty}
\newpage

\tableofcontents

\section{Introduction}
The direct CP-violation in $K\to\pi\pi$ decays,
represented by the ratio $\epe$,
 plays a  very important role in the tests of the Standard
Model (SM) and more recently in the tests of its possible extensions.
For recent reviews see \cite{Buras:2013ooa,Buras:2018wmb}. In fact
there are strong hints for  sizable new physics (NP) contributions
to $\epe$ from  Dual QCD  approach (DQCD)
\cite{Buras:2015xba,Buras:2016fys} that are supported to some extent by
RBC-UKQCD lattice collaboration \cite{Bai:2015nea,Blum:2015ywa}. Most recent
SM analyses at the NLO level can be found in  \cite{Buras:2015yba,Kitahara:2016nld} and a NNLO analysis is expected to appear soon \cite{Cerda-Sevilla:2016yzo}.
Most importantly, an improved result on $\epe$ from RBC-UKQCD lattice collaboration is expected this summer.

This situation motivated several authors
to look for various extensions of the SM which could bring the theory to agree with data \cite{Buras:2014sba,Buras:2015yca,Blanke:2015wba,Buras:2015kwd,Buras:2016dxz,Buras:2015jaq,Kitahara:2016otd,Endo:2016aws,Endo:2016tnu,
Cirigliano:2016yhc,Alioli:2017ces,Bobeth:2016llm,Bobeth:2017xry,Crivellin:2017gks,Bobeth:2017ecx,Endo:2017ums,Haba:2018byj,Chen:2018ytc,Chen:2018vog,Matsuzaki:2018jui,Haba:2018rzf}. In most of the models the rescue comes from the
modification of the Wilson coefficient of the dominant electroweak left-right (LR) penguin
operator $Q_8$, but also solutions through a modified contribution of the dominant
QCD LR penguin operator $Q_6$ could be considered  \cite{Buras:2015jaq}.

Here we want to emphasize that scalar-scalar and tensor-tensor four-fermion operators generated e.g. through
 tree-level exchanges of colour-singlet and colour-octet heavy mesons could also
give significant contributions to $\epe$ because they have, just like the $Q_6$ and $Q_8$ operators, chirally enhanced $K\to\pi\pi$  matrix elements. However, whereas in
the case of $Q_6$ and $Q_8$ operators significant progress in evaluating
their matrix elements relevant for $\epe$ by lattice QCD
has been made \cite{Bai:2015nea,Blum:2015ywa},
no lattice QCD calculations have been performed so far for these scalar-scalar and tensor-tensor operators although
their two-loop anomalous dimensions  have been known \cite{Ciuchini:1997bw,Buras:2000if} for almost two decades. In fact, to our knowledge, there exist
no analytic results for the matrix elements in question, even
 obtained by the vacuum insertion method  which, in any case, as
 already demonstrated in several studies,
totally misrepresents QCD.

We are aware of the fact that it will still take some time before lattice QCD
will be able to provide $K\to\pi\pi$ matrix elements for scalar-scalar and tensor-tensor operators. Yet, in view of the
hints for NP in $\epe$, we think it is time to estimate their matrix elements
in the framework of DQCD  \cite{Bardeen:1986vp,Bardeen:1986uz,Bardeen:1986vz,Bardeen:1987vg} which has been generalized in this decade
\cite{Buras:2014maa,Buras:2014apa,Buras:2015xba,Buras:2016fys} through the
inclusion of vector meson contributions and improved through a better matching
to short distance contributions. While
 not as precise as ultimate lattice QCD calculations, this approach offered over
 many years an insight in the lattice results and often, like was the case
of the $\Delta I=1/2$ rule \cite{Bardeen:1986vz}
and the parameter $\hat B_K$ \cite{Bardeen:1987vg}, provided results
 almost three decades before this was possible with lattice QCD.
The agreement between results from  DQCD  and lattice QCD is
remarkable, in particular considering the simplicity of the former approach
 compared to the very sophisticated and computationally demanding  numerical lattice QCD one. The most recent example of this agreement was an explanation by DQCD of the pattern of values of $\bsi$ and $\bei$ entering $\epe$ obtained by lattice QCD \cite{Buras:2015xba,Buras:2016fys} and of the pattern of lattice values for BSM  parameters $B_i$  in  $K^0-\bar K^0$  mixing \cite{Buras:2018lgu}. This is also the case for
 hadronic
matrix elements of the chromomagnetic operator presented recently in
\cite{Buras:2018evv} that are in agreement with the result from {the}
ETM collaboration \cite{Constantinou:2017sgv}.

Our paper is organized as follows. In Section~\ref{sec:3} we recall the
SM operators contributing to $K\to\pi\pi$ decays and construct a basis of four-quark BSM operators. Consisting exclusively of 13 products of colour singlet
scalar, vector and tensor bilinears, this complete basis is particularly useful for the calculations of hadronic  matrix elements in {the} DQCD approach and will thus be called {\em the  DQCD basis}.
In Section~\ref{sec:2} we recall very
briefly the elements of DQCD relevant for our paper. In Section~\ref{sec:5}
we perform the evolution of BSM operators from a very low factorization scale up
to scales $\mu=\ord(1\gev)$, the so-called meson evolution, in the chiral limit. While this is
a crude approximation, a recent analysis of BSM hadronic $K^0-\bar K^0$ matrix elements in this limit \cite{Buras:2018lgu} was able to explain at a semi-quantitative level the pattern of the values of these matrix elements obtained by
the ETM, SWME and RBC-UKQCD lattice QCD collaborations \cite{Carrasco:2015pra,Jang:2015sla,Garron:2016mva,Boyle:2017skn,Boyle:2017ssm}.

In Section~\ref{DQCDADM} we present the formulae for the quark-gluon evolution
in the DQCD basis while in Section~\ref{Matching} we demonstrate that the
patterns of meson evolution of BSM operators presented in Section~\ref{sec:5}
and of the quark-gluon evolution of Section~\ref{DQCDADM} are compatible with each other, assuring us that the matching of hadronic matrix elements evaluated
in DQCD and of their Wilson coefficients will be satisfactory. In Section~\ref{sec:BSM MEs} we calculate the matrix elements of all BSM operators at
leading order  in the DQCD
basis and, using the results for their meson evolution
of  Section~\ref{sec:5}, we obtain their values at {the} scale $\mu=\ord (1\gev)$.

In Section~\ref{sec:4} we introduce a different basis of 13 BSM operators which turns out to be particularly suited to the usual short distance (SD) QCD evolution. For this reason we call this basis
{\em the  SD basis}. We establish the relation between the DQCD  and SD bases
which, using the results of Section~\ref{sec:BSM MEs},
 allows us to obtain hadronic matrix elements of all BSM operators in the SD basis
at $\mu=\ord (1\gev)$. For completeness we give in Appendix~\ref{sec:BBM} also their values in the large $N$ limit.
The 13 BSM operators of a given chirality already mentioned  are all allowed by the $\text{SU(3)}_c\times \text{U(1)}_{\rm em}$ invariance. In Section~\ref{sec:SMEFT}, {following \cite{Aebischer:2018quc,Aebischer:2018csl}, we emphasize that in the SM effective field theory (SMEFT), based on the full SM gauge symmetry $\text{SU(3)}_c\times\text{SU(2)}_L\times \text{U(1)}_{Y}$,
only 7 four-quark BSM operators of a given chirality are allowed. We identify
these operators in the DQCD and SD bases.}

In Section~\ref{sec:6} we calculate the
$K\to\pi\pi$ matrix elements of  all BSM operators in two bases in question
for values of $\mu$ to be explored one day by lattice QCD. The results in the DQCD basis demonstrate once again that the pattern of meson evolution agrees with the one of SD evolution. The ones in the SD basis can now be used in the BSM phenomenology of $\epe$ and $\Delta I=1/2$ rule.

A brief summary of our results is given in  Section~\ref{sec:7}. In a number  of appendices we collect useful auxiliary material.
Phenomenological implications of our results will be presented elsewhere.

\boldmath
\section{$K\to\pi\pi$ Decays}\label{sec:3}
\unboldmath
\subsection{Preliminaries}
The isospin amplitudes $A_I$ in $K\to\pi\pi$
decays are introduced through
\begin{equation}\label{ISO1}
A(K^+\rightarrow\pi^+\pi^0)=\left(\frac{1}{h}\right)\left[\frac{3}{2} A_2 e^{i\delta_2}\right]~,
\end{equation}
\begin{equation}\label{ISO2}
A(K^0\rightarrow\pi^+\pi^-)=\left(\frac{1}{h}\right)\left[A_0 e^{i\delta_0}+ \sqrt{\frac{1}{2}}
 A_2 e^{i\delta_2}\right]~,
\end{equation}
\begin{equation}\label{ISO3}
A(K^0\rightarrow\pi^0\pi^0)=\left(\frac{1}{h}\right)\left[A_0 e^{i\delta_0}-\sqrt{2} A_2 e^{i\delta_2}\right]\,,
\end{equation}
\noindent
where the parameter $h$
 distinguishes between various normalizations of $A_{0,2}$ found in the literature. We use $h=1$ but the RBC-UKQCD collaboration uses $h=\sqrt{3/2}$
implying that their amplitudes  $A_{0,2}$ are by a factor $\sqrt{3/2}$ larger than ours. This difference cancels of course in all physical observables. All matrix elements listed below should be multiplied by $h$ in case $h=1$ is not used.
\subsection{SM Operators}
We begin by recalling the SM operators:
\\

{\bf Current-Current:}
\begin{equation}
  \label{eq:current-op}
\begin{aligned}
  Q_1 & = (\bar s d)_{V-A}(\bar uu)_{V-A} , &
  Q_2 & = (\bar s u)_{V-A}\,\,(\bar u d)_{V-A} ,
\end{aligned}
\end{equation}

{\bf QCD Penguins:}
\begin{equation}
  \label{eq:QCD-peng-op}
\begin{aligned}
  Q_3 & = (\bar s d)_{V-A}\!\!\sum_{q=u,d,s}(\bar qq)_{V-A} , &
  Q_4 & = (\bar s_{\alpha} d_{\beta})_{V-A}\!\!\sum_{q=u,d,s}(\bar q_{\beta} q_{\alpha})_{V-A} ,
\\
  Q_5 & = (\bar s d)_{V-A}\!\!\sum_{q=u,d,s}(\bar qq)_{V+A} , &
  Q_6 & = (\bar s_{\alpha} d_{\beta})_{V-A}\!\!\sum_{q=u,d,s}
      (\bar q_{\beta} q_{\alpha})_{V+A} ,
\end{aligned}
\end{equation}

{\bf Electroweak Penguins:}
\begin{equation}
  \label{eq:QED-peng-op}
\begin{aligned}
  Q_7 & = \frac{3}{2}\,(\bar s d)_{V-A}\!\!\sum_{q=u,d,s} e_q\,(\bar qq)_{V+A} , &
  Q_8 & = \frac{3}{2}\,(\bar s_{\alpha} d_{\beta})_{V-A}\!\!\sum_{q=u,d,s}
      e_q\,(\bar q_{\beta} q_{\alpha})_{V+A} ,
\\
  Q_9 & = \frac{3}{2}\,(\bar s d)_{V-A}\!\!\sum_{q=u,d,s}e_q\,(\bar q q)_{V-A} , &
  Q_{10} & =\frac{3}{2}\, (\bar s_{\alpha} d_{\beta})_{V-A}\!\!\sum_{q=u,d,s}e_q\,
       (\bar q_{\beta}q_{\alpha})_{V-A} .
\end{aligned}
\end{equation}
Here, $\alpha,\beta$ denote colour indices and $e_q$ the electric quark
charges reflecting the electroweak origin of $Q_7,\ldots,Q_{10}$. Finally,
$(\bar sd)_{V\pm A}\equiv \bar s_\alpha\gamma_\mu(1\pm\gamma_5) d_\alpha$.
As we are only interested in hadronic matrix elements in this paper,
the summations are only over $u,d,s$ quarks.

\subsection{BSM Operators}
\subsubsection{Chiral Fierz Identities}

The following 16 Dirac bilinears form the appropriate chiral basis for
a systematic classification of all $\Delta S=1$ weak operators beyond the Standard model (BSM):
\be
\{\Gamma^A\}=\{P_L,P_R,\gamma^\mu P_L,\gamma^\mu P_R,\sigma^{\mu\nu}\}, \qquad (A=1,..16)
\ee
with
\be
P_{L,R}=\f{1}{2}(1\mp\gamma_5), \qquad \{\gamma^\mu,\gamma^\nu\}=2 g^{\mu\nu}, \qquad \sigma^{\mu\nu}=\f{i}{2}[\gamma^\mu,\gamma^\nu]\,.
\ee

Within these conventions, the corresponding dual basis
\be
\{\Gamma_A\}=\{P_L,P_R,\gamma_\mu P_R,\gamma_\mu P_L,\f{1}{2}\sigma_{\mu\nu}\},
\ee
obeys the orthogonality property
\be
\text{Tr}(\Gamma_A\Gamma^B)=2 \delta^B_A\,.
\ee

If we substitute the matrix indices by parentheses $()$ and brackets $[\,]$ such
that each parenthesis/bracket represents a different spinor index, the completeness relation
\be
(1)[1]=\f{1}{2}(\Gamma_A][\Gamma^A),
\ee
leads then to the so-called chiral Fierz identities \cite{Nishi:2004st}
\be
(\Gamma^A)[\Gamma^B]=\f{1}{4}\text{Tr}(\Gamma^A\Gamma_C\Gamma^B\Gamma_D)(\Gamma^D][\Gamma^C).
\ee

Anticipating the fermion field anticommutation, we can turn all the four-quark operators into products of colour singlet bilinears according to four classes of Fierz identities:
\\[0.2cm]

{\bf Class A:}
\be\label{F1}
(\gamma^\mu P_L][\gamma_\mu P_L)=(\gamma^\mu P_L)[\gamma_\mu P_L],
\ee

{\bf Class B:}
\be\label{F2}
(\gamma^\mu P_L][\gamma_\mu P_R)=-2 (P_R)[P_L],
\ee

{\bf Class C:}
\be\label{F3}
(P_R][P_L)=-\f{1}{2}(\gamma^\mu P_L)[\gamma_\mu P_R],
\ee

{\bf Class D:}
\be\label{F4}
(P_L][P_L)=-\f{1}{2}(P_L)[P_L] -\f{1}{8}(\sigma^{\mu\nu} P_L)[\sigma_{\mu\nu} P_L],
\ee

\be\label{F5}
(\sigma^{\mu\nu}P_L][\sigma_{\mu\nu}P_L)=-6(P_L)[P_L] +\f{1}{2}(\sigma^{\mu\nu} P_L)[\sigma_{\mu\nu} P_L].
\ee

Mirror operators are obtained through an obvious chirality-flip
$(L\leftrightarrow R)$.
\boldmath
\subsubsection{Illustration with $\Delta S=2$ Operators}
\unboldmath

Tree-level neutral meson exchanges can lead to various BSM $\Delta S =2$ transitions. As a consequence, we have to consider the four classes of operators for a single set of Fierz-conjugate flavour indices $\{ab;cd\}=\{sd;sd\}$:
\begin{align}
 {\bf A}:& \quad(\bar s\gamma^{\mu}P_Ld][\bar s\gamma_{\mu}P_Ld)=(\bar s\gamma^{\mu}P_Ld)[\bar s\gamma_{\mu}P_Ld]=O_1,\label{eq:DS2A}\\
{\bf B}:& \quad(\bar s\gamma^{\mu}P_Ld][\bar s\gamma_{\mu}P_Rd)=-2(\bar sP_Rd)[\bar sP_Ld]=-2O_4, \\
{\bf C}:& \quad(\bar sP_Rd][\bar sP_Ld)=-\frac{1}{2}(\bar s\gamma^{\mu}P_Ld)[\bar s\gamma_{\mu}P_Rd]=O_5,\label{eq:DS2C}\\
{\bf D}:& \quad(\bar sP_Ld][\bar sP_Ld)=-\frac{1}{2}(\bar sP_Ld)[\bar sP_Ld]-\frac{1}{8}(\bar s\sigma_{\mu\nu}P_Ld)[\bar s\sigma^{\mu\nu}P_Ld]=O_3,\label{eq:DS2D1}\\
&\quad(\bar s\sigma_{\mu\nu}P_Ld][\bar s\sigma^{\mu\nu}P_Ld)=-6(\bar sP_Ld)[\bar sP_Ld]+\frac{1}{2}(\bar s\sigma_{\mu\nu}P_Ld)[\bar s\sigma^{\mu\nu}P_Ld]\notag \\
&\quad\qquad\qquad\qquad\qquad\,\,\,\,\,=-8O_2-4O_3.\label{eq:DS2D2}
\end{align}
Using this technology, we succeeded \cite{Buras:2000if,Buras:2018lgu} in turning the SUSY basis \cite{Gerard:1984bg,Gabbiani:1996hi}  represented by 1 SM operator $O_1$ and 4 BSM operators
$O_{2-5}$ into the SD basis of \cite{Buras:2000if}.

\boldmath
\subsubsection{Application to the SM $\Delta S=1$ Operators}\label{sec:7SM}
\unboldmath

In the case of tree-level charged or neutral meson exchanges leading to various
$\Delta S=1$ transitions we now have  to consider  four different sets of flavour indices $\{ab;cd\}$, namely
\be\label{eq:flavstruc}
\{su;ud\}\,,\qquad  \{sd;uu\}\,, \qquad  \{sd;dd\}\,,\qquad \{sd;ss\}\,,
\ee
 for  each class of operators. The last set of operators with a $(\bar s s)$
bilinear does not contribute to the $K\to\pi\pi$ matrix elements at the factorization scale, the starting point of our calculation. Yet, they may contribute above this scale through the evolution into other pure $\Delta I=1/2$ operators.

According to our generic Fierz classification and up to parity-transformations $(L\leftrightarrow R)$, this complete basis gives rise to $5\times 4=20$ linearly independent operators written as the product of two colour-singlet bilinears. But now, we want to build the optimal basis for a transparent evolution of the BSM operators orthogonal to the SM $Q_i$ displayed in  (\ref{eq:current-op})-(\ref{eq:QED-peng-op}). For that purpose let us first isolate the latter ones by selecting the appropriate combinations of operators in (\ref{eq:flavstruc}), with sums over $q=u,d,s$ understood.
\newline
\newline
{\bf Class  A:}
\be
(\bar s \gamma^\mu P_L u)[\bar u \gamma_\mu P_L d]=\f{1}{4} Q_2\,,\quad
(\bar s \gamma^\mu P_L d)[\bar u \gamma_\mu P_L u]=\f{1}{4} Q_1\,,
\ee
\be
(\bar s \gamma^\mu P_L d)[\bar q \gamma_\mu P_L q]=\f{1}{4} Q_3\,.
\ee
In fact this class covers {\it all} SM $(V-A)\times(V-A)$ operators since the charge matrix $e_q$ appearing in the electroweak weak penguins in (\ref{eq:QED-peng-op}) can be decomposed in the following way:
\be
\text{diag.}(2/3,-1/3,-1/3)=\text{diag.}(1,0,0)-(1/3)\text{diag.}(1,1,1)\,
\ee
 and implies
\begin{align}
Q_4   &= -Q_1+ Q_2  + Q_3   \,, \label{Q4a}\\
Q_9   &= \frac{3}{2}Q_1 -\frac{1}{2} Q_3  \, ,
\label{eq:5.12a} \\
Q_{10}   &= \frac{1}{2} Q_1+Q_2-\f{1}{2}Q_3 \,.
\label{eq:5.13a}
\end{align}
\newline
\newline
{\bf Class  B:}
\be
(\bar s P_R u)[\bar u P_L d]=-\f{1}{12}(Q_8+\frac{1}{2}Q_6)\,,\qquad
(\bar s P_R q)[\bar q P_L d]=-\f{1}{8} Q_6\,.
\ee
Here we use the fact that $\{ss;sd\}$ is the mirror partner of $\{sd;ss\}$.
\newline
\newline
{\bf Class  C:}
\be
(\bar s \gamma^\mu P_L u)[\bar u \gamma_\mu P_R d]= \f{1}{6} (Q_7+\frac{1}{2}Q_5)\,,\qquad
(\bar s \gamma^\mu P_L d)[\bar q \gamma_\mu P_R q]= \f{1}{4} Q_5\,.
\ee
So we end up with  a total of $7$ four-quark SM operators scattered in the first three classes, all of them being invariant under the discrete symmetry CPS which is the product of ordinary CP with $d\leftrightarrow s$ switch \cite{Bernard:1985wf}.

\boldmath
\subsubsection{Application to the BSM $\Delta S=1$ Operators}\label{sec:13BSM}
\unboldmath
As a consequence of our counting in the previous subsection, by orthogonality we are left with $20-7=13$ four-quark BSM operators linearly independent from the SM ones and violating CPS symmetry, namely

\vspace{2mm}
\noindent
- {\it one} {in \bf{class A}}:
\be\label{eq:classABSM}
A=(\bar s \gamma^\mu P_L d) [\bar d \gamma_\mu P_L d- \bar s \gamma_\mu P_L s]\,,
\ee
- {\it two} in {\bf{class B}}:
\be
B_1=(\bar s P_R d) [\bar u P_L u]\,, \qquad B_2=(\bar s P_R d) [\bar d P_L d]-(\bar s P_R s) [\bar s P_L d]\,,
\ee
- {\it two} in {\bf{class C}}:
\be
  C_1=(\bar s \gamma^\mu P_L u) [\bar u \gamma_\mu P_R d]\,, \quad  C_2=(\bar s \gamma^\mu P_L d) [\bar d \gamma_\mu P_R d-\bar s \gamma_\mu P_R s]\,,
\ee
- {\it eight} in {\bf{class D}}:
\be\label{eq:XSLLqd}
 D_1=(\bar s P_L u) [\bar u P_L d]\,, \qquad  D_2=(\bar s  P_L d) [\bar u  P_L u]\,,
\ee

\be
 D_3=(\bar s P_L d) [\bar d P_L d]\,, \qquad  D_4=(\bar s P_L d) [\bar s P_L s]\,,
\ee

\be
 D_1^*=-(\bar s \sigma^{\mu\nu} P_L u) [\bar u  \sigma_{\mu\nu}P_L d]\,, \qquad  D_2^*=-(\bar s  \sigma^{\mu\nu} P_L d) [\bar u  \sigma_{\mu\nu}  P_L u]\,,
\ee

\be
 D_3^*=-(\bar s  \sigma^{\mu\nu} P_L d) [\bar d  \sigma_{\mu\nu}P_L d]\,, \qquad  D_4^*=-(\bar s  \sigma^{\mu\nu}P_L d) [\bar s  \sigma_{\mu\nu} P_L s]\,.
\ee

Finally we want to emphasize the following virtue of the chosen basis of BSM
operators in
which, in contrast to the SM basis in (\ref{eq:QCD-peng-op}) and
 (\ref{eq:QED-peng-op}), no summation over quarks is performed. Indeed when considering various extensions of the SM it often turns out that
a new heavy mediator, vector or scalar, couples at tree level only to up-quarks or down-quarks  but not to both. Moreover right-handed  up- and  down-quark couplings to new mediators, not related by $\text{SU(2)}_L$ symmetry,  could
differ by much from each other. In this manner this basis, to be called
DQCD basis, is not only appropriate to the evaluation of hadronic matrix elements
but also useful for model building.

\section{Dual QCD basics}\label{sec:2}
The explicit calculation of the contributions of pseudoscalars to hadronic matrix elements of local operators is based on  a truncated chiral Lagrangian describing the low energy
interactions of the lightest mesons \cite{Chivukula:1986du,Bardeen:1986vp,Bardeen:1986uz}
\be\label{chL}
L_{tr}=\frac{F^2}{8}\left[\text{Tr}(D^\mu UD_\mu U^\dagger)+r\text{Tr}(mU^\dagger+\text{h.c.})-\frac{r}{\Lambda^2_\chi}\text{Tr}(mD^2U^\dagger+\text{h.c.})\right] {\,,}
\ee
where
\be\label{UU}
U=\exp(i\sqrt{2}\frac{\Pi}{F}), \qquad
\Pi=\sum_{\alpha=1}^8\lambda_\alpha\pi^\alpha{\,,}
\ee
is the unitary chiral matrix describing the octet of light pseudoscalars and transforming as $U\rightarrow g_LUg_R^{\dag}$ under the chiral symmetry $SU(3)_L\times SU(3)_R$.
The parameter $F$ is related to  the weak decay constants $F_\pi\approx 130\mev$
and $F_K\approx 156\mev$ through
\be\label{FpiFK}
F_\pi=F\left(1+\frac{m_\pi^2}{\Lambda^2_\chi}\right), \qquad F_K=F\left(1+\frac{m_K^2}{\Lambda^2_\chi}\right),
\ee
so that $\Lambda_\chi\approx 1.1\gev$.
The diagonal mass matrix $m$ involving $m_u$, $m_d$ and $m_s$ is such
that
\be\label{rr}
r(\mu)=\frac{2 m_K^2}{m_s(\mu)+m_d(\mu)},
\ee
with $r(1\gev)\approx 3.75\gev$ for $(m_s+m_d)(1\gev)\approx 132\mev$. Compared to large $N$ Chiral Perturbation Theory, there is a one-to-one correspondence to the low energy parameters introduced in \cite{Gasser:1983yg,Gasser:1984gg}
\be
\Lambda^2_\chi=\frac{f^2}{8L_5}, \qquad r=2B_0\,.
\ee

The flavour-singlet $\eta_0$ meson decouples due to the large mass $m_0$ generated
by the non-perturbative $U(1)_A$ anomaly. Consequently the matrix $\Pi$ in (\ref{UU}) reads

\begin{align}\label{eq:Umatrix}
 \Pi = \begin{pmatrix}
                \pi^0+\frac{1}{\sqrt{3}}\eta_8 & \sqrt{2}\pi^+ & \sqrt{2}K^+ \\
		\sqrt{2}\pi^- & -\pi^0+\frac{1}{\sqrt{3}}\eta_8 &  \sqrt{2}K^0 \\
	 \sqrt{2}K^- &  \sqrt{2}\bar K^0 & -\frac{2}{\sqrt{3}}\eta_8
                \end{pmatrix}.
\end{align}
In order to calculate the matrix elements of the local operators in question we
need meson representations of colour-singlet quark bilinears. Only currents and densities are directly extracted from the effective Lagrangian in (\ref{chL}). They
are given respectively as follows
\be\label{VAc}
\bar q^b_L\gamma_\mu q^a_L=i\frac{F^2}{8}\left\{(\partial_\mu U)U^\dagger-U(\partial_\mu U^\dagger)+
\frac{r}{\Lambda^2_\chi}\left[(\partial_\mu U)m^\dagger-m(\partial_\mu U^\dagger)\right]\right\}^{ab},
\ee
\be\label{RLd}
\bar q_R^b q_L^a=-\frac{F^2}{8}r\left[U-\frac{1}{\Lambda_\chi^2}\partial^2U\right]^{ab}\,,
\ee
with $U$ turned into $U^\dagger$ under parity. As a matter of fact, the chiral correction to densities is meaningful for the $Q_6$ operator only, as seen in eq.~(\ref{eq:Q60}). For $Q_8$ and the other density-density operators, $\mathcal{O}(p^4)$ mass terms with additional low-energy constants should be introduced in (\ref{chL}). For the sake of consistency, we will thus work in the chiral limit for all of them.

The lowest-order chiral realization of tensor bilinears requires two derivatives to get the correct Lorentz structure. It thus involves yet another dimensionful low-energy constant \cite{Mertens:2011ts}:
\begin{equation}
  \bar q_R^b \sigma_{\mu\nu} q^a_L = -i\frac{F^2}{4\Lambda_{\chi}'}[\partial_{\mu}U\partial_{\nu}U^{\dag}U-\partial_{\nu}U\partial_{\mu}U^{\dag}U]^{ab}\,.
\end{equation}
In the large $N$ limit, all four-quark operators factorize into two colour singlet bilinears. As a consequence, at $\mathcal{O}(p^2)$ the only relevant colour-singlet bilinears are
\be
(\gamma^\mu P_L)^{ba}=i \f{F^2}{4} (\partial^\mu U U^\dagger)^{ab}, \qquad (P_L)^{ba}=- \f{F^2}{8} r (U)^{ab}\,,
\ee

and their parity partners, with
\begin{align}
{\bf Class~A}&:  \quad-\left[\f{F^2}{4}\right]^2 (\partial^\mu U U^\dagger)^{ab}
 (\partial_\mu U U^\dagger)^{cd}\label{N1}{\,,}
\\
{\bf Class~B}&:  \quad+\left[\f{F^2}{8}\right]^2 r^2(U^\dagger)^{ab}
 (U)^{cd}{\,,}
\\
{\bf Class~C}&:  \quad-\left[\f{F^2}{4}\right]^2 (\partial^\mu U U^\dagger)^{ab}
 (\partial_\mu U^\dagger U)^{cd}{\,,}
\\
{\bf Class~D}&:  \quad+\left[\f{F^2}{8}\right]^2 r^2(U)^{ab}
 (U)^{cd}\,.\label{N4}
\end{align}

\section{Meson evolution in the  DQCD basis}\label{sec:5}
The  formulae (\ref{N1})-(\ref{N4}) apply to the strict large $N$ limit
at which the factorization of matrix elements is valid. In this limit it
is not possible to determine the scale associated to these matrix elements.
To this end one has to calculate non-factorizable contributions represented by
loops in the meson theory. The factorization scale is then the scale at which
these non-factorizable contribution vanish. Quite generally the factorization scale is found to be at very low scales $\ord(m_\pi)$ and in
order to obtain the matrix elements at scales $\ord(1\gev)$ one has to evolve them with the help of the meson evolution. As in our paper this evolution will be
performed in the chiral limit, the factorization scale is simply at zero momentum.

Like in our recent paper on BSM hadronic matrix elements for $K^0-\bar K^0$
mixing \cite{Buras:2018lgu}, we will work in the $\mathcal{O}(p^2)$ chiral limit with
\begin{equation}
  \frac{p^2}{(4\pi F)^2} = \mathcal{O}(1/N)\,.
\end{equation}

While this is a rough
approximation, it has been demonstrated there that
 the pattern of matrix elements evaluated in DQCD at a scale $\Lambda=(0.65\pm0.05)\gev$
agrees well with the pattern  at $\mu=1\gev$
obtained from lattice QCD results  at $\mu=3\gev$ through
 the usual perturbative quark QCD evolution with
\begin{equation}
  \frac{\alpha_s}{4\pi} = \mathcal{O}(1/N)\,.
\end{equation}
  This makes us confident that
the results presented here on the basis of the \underline{same} $1/N$ counting and the \underline{same} $\Lambda$-to-$\mu$ identification have rather similar uncertainties.

Let us first emphasize that all $K\rightarrow\pi\pi$ matrix elements of chirally-flipped (relative to the SM operators)  mirror operators denoted by a prime can be obtained from the results of RBC-UKQCD collaboration
by just {\it reversing their signs}.
What remains for us is the
calculation of the matrix elements which cannot be expressed in
terms of SM ones.

The flavour $SU(n)$ generators already introduced in (\ref{UU}) for $n=3$ are normalized such that
\be
\text{Tr}\left(\lambda_{\alpha}\lambda_\beta\right)=2\delta_{\alpha\beta}.
\ee
If we again substitute the matrix indices by parenthesis () and brackets [ ] such that each parenthesis/bracket represents now a different flavour index, then the completeness relation among matrices of the fundamental representation of $SU(n)$
\be\label{eq:lambda2}
\frac{1}{2}(\lambda_{\alpha})^{ab}(\lambda_{\alpha})^{cd}+\frac{1}{n}\delta^{ab}\delta^{cd}=\delta^{ad}\delta^{cb},
\ee
simply reads
\be
\frac{1}{2}(\lambda_{\alpha})\left[\lambda_{\alpha}\right]+\frac{1}{n}(\mathbb{1})[\mathbb{1}]=(\mathbb{1}][\mathbb{1})\,.
\ee
With the background field technology used in \cite{Fatelo:1994qh}, the relevant {one-loop} operator evolutions from the factorization scale taken at zero momentum (denoted 0) to the cut-off-momentum (denoted $\Lambda$) can also be classified according to these identities:

\begin{align}
 \text{\bf{Class A}}&:(\partial^{\mu} UU^{\dag})^{ab}(\partial_{\mu} UU^{\dag})^{cd}(\Lambda)=(\partial^{\mu} UU^{\dag})^{ab}(\partial_{\mu} UU^{\dag})^{cd}(0)\label{MEA} \\\notag
 &- 4\,\left(\frac{\Lambda}{4\pi F}\right)^2\left[(\partial^{\mu} UU^{\dag})^{ad}(\partial_{\mu} UU^{\dag})^{cb} +\frac{1}{2}\delta^{ad}(\partial^{\mu}U\partial_{\mu}U^{\dagger})^{cb} \right.
 \\\notag
 &\left.\qquad\qquad\qquad\quad+\frac{1}{2}\delta^{cb}(\partial^{\mu}U\partial_{\mu}U^{\dagger})^{ad}\right](0),
 \\
 \text{\bf{Class B}}&:
 (U^{\dagger})^{ab}(U)^{cd}(\Lambda)=(U^{\dagger})^{ab}(U)^{cd}(0)\label{MEB}\\\notag
 &\quad\quad\quad\quad\quad\quad\quad\quad+ 4\,\left(\frac{\Lambda}{4\pi F}\right)^2\left[(U^{\dagger}U)^{ad}\delta^{cb}-\frac{1}{n}(U^{\dagger})^{ab}(U)^{cd}\right](0),
 \\
  \text{\bf{Class C}}&:(\partial^{\mu} UU^{\dag})^{ab}(\partial_{\mu} U^{\dag}U)^{cd}(\Lambda)=(\partial^{\mu} UU^{\dag})^{ab}(\partial_{\mu} U^{\dag}U)^{cd}(0)\label{MEC}\\\notag
  &\quad\quad\quad\quad\quad\quad\quad\quad+ 4\,\left(\frac{\Lambda}{4\pi F}\right)^2M^2\left[(U)^{ad}(U^{\dagger})^{cb}-\frac{1}{n}\delta^{ab}(U^{\dagger}U)^{cd}\right](0),\\
 \text{\bf{Class D}}&:(U)^{ab}(U)^{cd}(\Lambda)=(U)^{ab}(U)^{cd}(0) \\\notag
 &\quad\qquad\qquad\qquad- 4\,\left(\frac{\Lambda}{4\pi F}\right)^2\left[(U)^{ad}(U)^{cb}-\frac{1}{n}(U)^{ab}(U)^{cd}\right](0)\label{MED},
\end{align}
with $M^2$, an $SU(3)$-breaking mass term equal to $m_K^2$ for $\Delta S=2$ transitions and of order $(m_K^2-m_{\pi}^2)$ for $\Delta S=1$ ones. Had we worked with a nonet of light pseudo-scalars, the $1/n$ term resulting from the purely non-perturbative axial anomaly would have been absent $(m_0=0)$.

 As already mentioned below (\ref{eq:flavstruc}), the set of operators $\{sd;ss\}$ may
evolve into other $\Delta I=1/2$ ones. Such is indeed the case in the meson evolution where only the one in Class A  induces the SM $Q_4$ operator.
We also observe a reordering of the flavour  indices with mixing between operators of classes B and C. Eventually, the mirror operators are again obtained through a parity transformation ($U\leftrightarrow U^{\dagger}$).

Applied to the 13 BSM operators classified in section \ref{sec:13BSM}, these non-factorizable meson evolutions imply:
\\

{\bf Class A}: if $A\equiv \{sd;dd\}-\{sd;ss\}$
\begin{equation}\label{eq:ALambda}
  A(\Lambda)=\left[1-4\left(\frac{\Lambda}{4\pi F}\right)^2\right]A(0)\,,
\end{equation}

{\bf Class B}: if $B_1 \equiv \{sd;uu\}$ and $B_2 \equiv \{sd;dd\}-\{ss;sd\}$ with $\{ss;sd\}=\{sd;ss\}'$
\begin{equation}
  B_{1,2}(\Lambda)=\left[1-\frac{4}{n}\left(\frac{\Lambda}{4\pi F}\right)^2\right]B_{1,2}(0)\,,
\end{equation}

{\bf Class C}: if $C_1 \equiv \{su;ud\}$ and $C_2 \equiv \{sd;dd\}-\{sd;ss\}$
\begin{equation}
  C_{1,2}(\Lambda)=C_{1,2}(0)-16\frac{M^2}{r^2}\left(\frac{\Lambda}{4\pi F}\right)^2B_{1,2}(0)\,,
\end{equation}

{\bf Class D}: if $D_1 \equiv \{su;ud\}$, $D_2 \equiv \{sd;uu\}$, $D_3 \equiv \{sd;dd\}$, $D_4 \equiv \{sd;ss\}$,
\begin{align}
D_{1}(\Lambda) &=\left[1+\frac{4}{n}\left(\frac{\Lambda}{4\pi \label{eq:D1Lambda} F}\right)^2\right]D_{1}(0)-4\left(\frac{\Lambda}{4\pi F}\right)^2D_2(0)\,, \\
  D_{2}(\Lambda) &=\left[1+\frac{4}{n}\left(\frac{\Lambda}{4\pi F}\right)^2\right]D_{2}(0)-4\left(\frac{\Lambda}{4\pi F}\right)^2D_1(0)\,, \\
D_{3}(\Lambda) &=\left[1+(\frac{4}{n}-4)\left(\frac{\Lambda}{4\pi F}\right)^2\right]D_{3}(0)\,, \\
    D_{4}(\Lambda) &=\left[1+(\frac{4}{n}-4)\left(\frac{\Lambda}{4\pi F}\right)^2\right]D_{4}(0)\,.\label{eq:D4Lambda}
\end{align}

They agree respectively with the evolution derived in \cite{Buras:2018lgu} for the $O_{1,4,5}\,\,\,\{sd;sd\}$ operators defined in (\ref{eq:DS2A})-(\ref{eq:DS2C}) and
for the $O_2\,\,\,\{sd;sd\}$ operator extracted from (\ref{eq:DS2D1})-(\ref{eq:DS2D2}). However, to infer the meson evolution of the four tensor-tensor operators $D_i^{*}$ above the factorization scale, we now have to rely on the SD running pattern.

\section{Quark-gluon evolution in DQCD basis}\label{DQCDADM}
In order to study the short distance RG evolution from $\mu= 1 \gev$  to higher scales we collect here
the fundamental equations in the leading order approximation. To see
the pattern of this evolution we keep first only the
 first leading logarithms. One has then for $\mu_2>\mu_1$
\be\label{RGOi}
\langle {X}_i(\mu_2)\rangle =
\langle {X}_i(\mu_1)\rangle \left(1-\aspi {\hat{\gamma}^{(0)}_{ii}}\ln\left(\f{\mu_2}{\mu_1}\right)\right) -
 \langle {X}_j(\mu_1)\rangle \aspi {\hat{\gamma}^{(0)}_{ij}}\ln\left(\f{\mu_2}{\mu_1}\right)\,,
\ee
where ${X}_i$ denote generically the operators in the DQCD basis.
We group them in classes I-IV with no mixing under renormalization between various classes. This will also allow us a transparent comparison with the so-called SD basis that
we discuss in Section~\ref{sec:4}.

The anomalous dimension matrix (ADM) for all SD operators is then given in the DQCD basis  as follows  (in units of $\as/4\pi$):
\\

{\bf Class I}

\be\label{eq:ClI}
\hat{\gamma}^{(0)}( A) = 4\,,
\ee

{\bf Class II}

\bea\label{ADMII}
\hat{\gamma}^{(0)}(B_1, C_1) &=& \left(
\begin{array}{cc}
 -6 N+\frac{6}{N} & 0 \\
 12 & \frac{6}{N} \\
\end{array}
\right)=\left(
\begin{array}{cc}
 -16 & \,\,\,0 \\
 12 & \,\,\,2 \\
\end{array}
\right){\,,}
\eea
with the same matrix for the operators $B_2,C_2$,
\\

{\bf Class III}
\newline
\newline
\bea\label{eq:CIII}
\hat{\gamma}^{(0)}(D_1,D_2,D_1^*,D_2^*) &=& \left(
\begin{array}{cccc}
 -6 N+\frac{6}{N} & 6 &
   -\frac{1}{N} & \frac{1}{2} \\
 6 & -6 N+\frac{6}{N} &
   \frac{1}{2} & -\frac{1}{N} \\
 -\frac{48}{N} & -24 & 2
   N-\frac{2}{N} & 6 \\
 -24 & -\frac{48}{N} & 6 & 2
   N-\frac{2}{N} \\
\end{array}
\right)\\\notag
&=&\left(
\begin{array}{cccc}
 -16 & 6 & -\frac{1}{3} & \frac{1}{2} \\
 6 & -16 & \frac{1}{2} & -\frac{1}{3} \\
 -16 & -24 & \frac{16}{3} & 6 \\
 -24 & -16 & 6 & \frac{16}{3} \\
\end{array}
\right)\,,
\eea

{\bf Class IV}

\bea\label{eq:ClIV}
\hat{\gamma}^{(0)}(D_3,D_3^*) &=& \left(
\begin{array}{cc}
 -6 N+\frac{6}{N}+6 &
   \frac{1}{2}-\frac{1}{N} \\
 -24-\frac{48}{N} & 2
   N-\frac{2}{N}+6 \\
\end{array}
\right)=\left(
\begin{array}{cc}
 -10 & \,\,\,\frac{1}{6} \\
 -40 & \,\,\,\frac{34}{3} \\
\end{array}
\right)\,,
\eea
with the same matrix for the operators $D_4,D_4^*$.

The numerical values of the elements of these ADMs correspond
to $N=3$ but their explicit $N$ dependence will turn out to be very useful soon.

Of particular interest here are the large entries in the elements $(3,2)$, and $(4,1)$ in (\ref{eq:CIII}) and $(2,1)$ in (\ref{eq:ClIV}) that
imply
large mixing of the scalar-scalar operators into tensor-tensor operators.
We will see soon that this feature enhances the matrix elements of tensor-tensor
operators in the process of $\mathcal{O}(1/N)$ meson evolution. This feature has
some analogy  to the observation made in
 \cite{Aebischer:2017gaw,Gonzalez-Alonso:2017iyc} where the  QED short distance RG evolution of NP contributions to charged-current induced leptonic
and semileptonic meson decays has been presented, focusing on chirality-flipped  operators at the quark level. It has been pointed out that the large mixing of the tensor-tensor operators into the scalar-scalar ones has an important impact on the phenomenology. Recently this aspect has also been discussed in
the context of $R(D^{(*)})$ anomalies in
\cite{Feruglio:2018fxo,Becirevic:2018afm}.

In fact the one-loop QED diagrams responsible for this mixing  are the same as the one-loop  QCD diagrams with gluon replaced by photon, QCD coupling
replaced by QED one and colour matrices replaced by charge ones. Even if the RG evolution
of the QED coupling constant is different from the QCD one, the pattern of mixing
analysed in \cite{Aebischer:2017gaw,Gonzalez-Alonso:2017iyc} is very similar
to the one in  (\ref{eq:CIII})\footnote{See lower right corner of $\gamma^T_{em}$ in (2.3) of \cite{Gonzalez-Alonso:2017iyc}, compared to $D_2-D_1^*$ central submatrix in (\ref{eq:CIII}).}.

The reason why in  \cite{Aebischer:2017gaw,Gonzalez-Alonso:2017iyc}  tensor operators have the impact on the scalar ones, as opposed to the case discussed by us, is simply related to the known fact that while the evolution of the matrix elements of operators is governed by the ADM of operators, the evolution of their Wilson coefficients, analysed in  \cite{Aebischer:2017gaw,Gonzalez-Alonso:2017iyc}, is governed  by the corresponding transposed matrix.

While in \cite{Aebischer:2017gaw,Gonzalez-Alonso:2017iyc,Feruglio:2018fxo,Becirevic:2018afm} the large mixing in question had impact on the phenomenology of $B$-meson decays, in our case it will
have significant impact on $\epe$.

\section{Matching SD-LD evolutions in DQCD basis}\label{Matching}
From (\ref{eq:ClI})-(\ref{eq:ClIV}), the short distance quark-gluon non-factorizable evolutions read
\begin{align}
  A(\mu_2) & = \left[1-4\,\frac{\alpha_s}{4\pi}\ln\left(\frac{\mu_2}{\mu_1}\right) \right]A(\mu_1) \,,\\
  B_{1,2}(\mu_2) & = B_{1,2}(\mu_1)\,,\\
    C_{1,2}(\mu_2) & = C_{1,2}(\mu_1)-12\frac{\alpha_s}{4\pi}\ln\left(\frac{\mu_2}{\mu_1}\right) B_{1,2}(\mu_1) \,,\\
    D_{1,2}(\mu_2) & = \label{eq:D12} D_{1,2}(\mu_1)-6\frac{\alpha_s}{4\pi}\ln\left(\frac{\mu_2}{\mu_1}\right) \left[D_{2,1}+\frac{1}{12}D^*_{2,1}\right](\mu_1) \,,\\
      D_{3,4}(\mu_2) & = \left[1-6\,\frac{\alpha_s}{4\pi}\ln\left(\frac{\mu_2}{\mu_1}\right) \right]D_{3,4}(\mu_1)-\frac{1}{2}\,\frac{\alpha_s}{4\pi}\ln\left(\frac{\mu_2}{\mu_1}\right)D^*_{3,4}(\mu_1) \,,\\
      D_{1,2}^*(\mu_2) & = D_{1,2}^*(\mu_1)+ \label{eq:Dst12} 24\frac{\alpha_s}{4\pi}\ln\left(\frac{\mu_2}{\mu_1}\right) \left[D_{2,1}-\frac{1}{4}D^*_{2,1}\right](\mu_1) \,,\\
  D_{3,4}^*(\mu_2) & = D_{3,4}^*(\mu_1)+ 24\frac{\alpha_s}{4\pi}\ln\left(\frac{\mu_2}{\mu_1}\right) \left[D_{3,4}-\frac{1}{4}D^*_{3,4}\right](\mu_1)\label{eq:Dst34} \,,
\end{align}
\\

by
\begin{itemize}
\item subtracting the  $(-6N+6/N)$ and $(2N-2/N)$ diagonal contributions to scalar-scalar and tensor-tensor operators, given the anomalous dimension of the $\bar qq$ and $\bar q\sigma^{\mu\nu}q$ bilinears \cite{Broadhurst:1994se,Bobeth:2007dw}:
\begin{equation}\label{AST}
  \gamma_{S}=-3\frac{N^2-1}{N},\quad \gamma_{T}=\frac{N^2-1}{N},
\end{equation}

  \item dropping the subleading $1/N$ terms in the ADMs (\ref{ADMII})-(\ref{eq:ClIV}):
  \begin{equation}
    \frac{\alpha_s}{N}=\mathcal{O}(\frac{1}{N^2})\,.
  \end{equation}
\end{itemize}

From (\ref{eq:ALambda})-(\ref{eq:D4Lambda}), these non-factorizable evolutions are compatible with the long-distance ones in the nonet approximation $(m_0\rightarrow 0)$ with
\begin{equation}
  \left(\frac{r(\mu_2)}{r(\mu_1)}\right)^2=1+2\gamma_m\,\frac{\alpha_s}{4\pi}\ln\left(\frac{\mu_2}{\mu_1}\right),\quad \gamma_m=-\gamma_S,
\end{equation}
since the tensor-tensor operators vanish at the factorization scale at $\mathcal{O}(p^2)$. Comforted by such a consistent matching of mixing pattern, we extend the octet meson evolution $(m_0\rightarrow \infty)$ to the non-factorizable tensor-tensor operators as follows:
\begin{align}
  D^*_{1,2}(\Lambda) &=+16\left(\frac{\Lambda}{4\pi F}\right)^2D_{2,1}(0)\,,\label{D*D12} \\
  D^*_{3,4}(\Lambda) &=+\frac{32}{3}\left(\frac{\Lambda}{4\pi F}\right)^2D_{3,4}(0)\,,\label{D*D34}
\end{align}
through the relative $(-4)$ factor between the $D$-to-$D$ and $D^*$-to-$D$ SD evolutions (\ref{eq:D12})-(\ref{eq:Dst34}) taken over to the $D$-to-$D$ LD evolutions (\ref{eq:D1Lambda})-(\ref{eq:D4Lambda}). They agree with the evolution derived in \cite{Buras:2018lgu} for the $O_3\,\,\,\{sd;sd\}$ operator defined in (\ref{eq:DS2D1}).

\section{BSM matrix elements in DQCD basis}\label{sec:BSM MEs}

\boldmath
\subsection{Large $N$ limit}
\unboldmath
Having established the meson evolution from the factorization scale
(corresponding in the chiral limit to $\Lambda=0$) to $\Lambda=\ord(1\gev)$, what remains to be done is the calculation of the matrix elements of all 13  four-quark BSM operators in the large $N$ limit that here will be generically denoted by
\be
\langle X_i(0) \rangle_I \equiv
\langle \left(\pi\pi\right)_I \left| X_i(0) \right| K \rangle \, ,
\label{eq:5.1}
\ee
with $I=0,2$ being strong isospin and $X_i=A, B_{1,2},...$.

When calculating the non-zero matrix elements of the BSM operators in Class B and D, we have to take into account the fact that the partial $\langle \pi\pi|U_{ds}|K^0\rangle$ contribution to on-shell $K\rightarrow \pi\pi$ decay amplitudes is precisely canceled by a non-local pole diagram involving the strong $K^0\rightarrow\pi\pi\bar{K}^0$ vertex followed by the $\bar K^0$ annihilation into the vacuum through the non-vanishing $\langle \bar{K}^0|U_{ds}|0\rangle$ weak matrix element. Considering charge conservation and Lorentz invariance, the recipe is to simply neglect any contribution from the identity when expanding the $U_{uu,dd,ss}$ components of density-density operators in Classes B and D. In giving the values of the matrix elements,
we drop the overall $+i$ factor that is immaterial for physical applications. In the large $N$ limit, the non-vanishing BSM matrix elements in the DQCD basis are then given as follows ($h=1$):
\begin{align}
\langle A(0)\rangle_0=& +\frac{F}{12} (m_K^2-m_\pi^2), & \langle A(0)\rangle_2= &-\sqrt{2}\frac{F}{12} (m_K^2-m_\pi^2)\,,~~~~~~~ \label{AA}\\
\langle B_1(0)\rangle_0=& +\frac{F}{12} r^2, & \langle B_1(0)\rangle_2=& \,-\frac{F}{24\sqrt{2}}r^2\,,~~~~~~~ \\
\langle B_2(0)\rangle_0=& +\frac{F}{24} r^2, & \langle B_2(0)\rangle_2=& \,+\frac{F}{24\sqrt{2}}r^2\,,~~~~~~~ \\
\langle C_1(0)\rangle_0= & {{-}}\frac{F}{6}(m_K^2-m_\pi^2),& \langle C_1(0)\rangle_2=& \, {{-}}\frac{F}{6\sqrt{2}}(m_K^2-m_\pi^2)\,,~~~~~~~ \\
\langle C_2(0)\rangle_0= & {-}\frac{F}{12}(m_K^2-m_\pi^2),& \langle C_2(0)\rangle_2=& \, {+}\sqrt{2}\frac{F}{12}(m_K^2-m_\pi^2)\,,~~~~~~~
\end{align}

\begin{align}
\langle D_1(0)\rangle_0=& - \frac{F}{24} r^2, &\langle D_1(0)\rangle_2=& - \frac{F}{24\sqrt{2}} r^2\,,\\
\langle D_2(0)\rangle_0=&- \frac{F}{24} r^2, & \langle D_2(0)\rangle_2=& \,- \frac{F}{24\sqrt{2}} r^2\,,~~~~~~~
\\
\langle D_3(0)\rangle_0=&- \frac{F}{12} r^2, & \langle D_3(0)\rangle_2=&  \, +\frac{F}{24\sqrt{2}} r^2\,,~~~~~~~
\end{align}
with the chiral enhancement factor $r^2 \gg (m_K^2-m_{\pi}^2)$ as seen from (\ref{rr}) and $F\sim F_{\pi}$ as seen from (\ref{FpiFK}).  Those of the mirror operators differ by sign only.

The matrix elements for the operator $D_4$, containing three s-quarks, as well as for the tensor-tensor operators $D_{1,2,3,4}^*$ involving at least four derivatives vanish:

\be\label{T0}
\langle D_4(0)\rangle_I= \langle D_1^*(0)\rangle_I=\langle D_2^*(0)\rangle_I=\langle D_3^*(0)\rangle_I=\langle D_4^*(0)\rangle_I=0\,.
\ee

We are thus left with 8 four-quark BSM matrix elements for a given chirality and isospin, each of them expressed in terms of either $(m_K^2-m^2_\pi)$ or $r^2(\mu)$ in the chiral limit considered.

\subsection{Summary of hadronic matrix calculations}
We have now completed the calculation of 13 BSM hadronic matrix elements
evaluated at the cut-off scale $\Lambda$, which is governed by the
general formula
\be\label{FME}
\langle X_i(\Lambda)\rangle_I=\left[\delta_{ij}+a_{ij}\left(\frac{\Lambda}{4\pi F}\right)^2\right] \langle X_j(0)\rangle_I
\ee
with  the coefficients $a_{ij}$ to be extracted from
(\ref{eq:ALambda})-(\ref{eq:D4Lambda}), (\ref{D*D12}) and (\ref{D*D34}) and
with the matrix elements $\langle X_j(0) \rangle_I$ collected above. In evaluating these matrix elements one should set $n=3$.

 In the numerical evaluation of matrix elements we will deal with two scales, $\Lambda$ explicitly seen in (\ref{FME}) and $\mu$ in $r(\mu)$ hidden in
$\langle X_j(0)\rangle_I$. In this context it is useful to make the following
comments:

\begin{itemize}
\item
The $\Lambda$ dependence is present only in the non-factorizable part of the
matrix elements as given above. The scale $\mu$ present in $r(\mu)$ is at this stage not related to $\Lambda$.
\item
As far as meson evolution in the chiral limit
is concerned, there is no distinction between the matrix elements $\langle {X}_i\rangle_0$ and $\langle {X}_i\rangle_2$ so that this distinction is
fully described by the values of these matrix elements in the large $N$ limit,
that is $\langle X_j(0)\rangle_I$.
\end{itemize}

Concerning the value of $\Lambda$ we will set it at $0.7\gev$. Evaluating then
$r(\mu)$ at  $\mu=1\gev$, we will interpret the resulting values of matrix elements  as valid at $\mu=1\gev$. From there on we will use the standard renormalization group evolution as summarized in Section~\ref{DQCDADM}, thereby
summing this time leading logarithms to all orders of perturbation
theory.
\subsection{Number of BSM matrix elements to be calculated}\label{Counting}
In principle one has to evaluate 13 matrix elements  for a given
chirality and isospin at the factorization scale. They are given in
(\ref{AA})-(\ref{T0}). However by definition $\bar s s$ bilinears do not
contribute to the $K\to\pi\pi$ matrix elements at the factorization scale.
Consequently  the matrix elements of $D_4$ and $D_4^*$  vanish, while the matrix elements of $A$ and  $B_2$ and $C_2$ can then be expressed in terms of the SM ones as follows
\be
\langle A\rangle_I= \frac{1}{6} (\langle Q_3\rangle_I-\langle Q_9\rangle_I),
\ee
\be
\langle B_2\rangle_I= -\frac{1}{12} (\langle Q_6\rangle_I-\langle Q_8\rangle_I),
\ee
\be
\langle C_2 \rangle_I= \frac{1}{6} (\langle Q_5\rangle_I-\langle Q_7\rangle_I)\,,
\ee
wherever $\langle \bar s \Gamma_A s\rangle=0$. Therefore,  in the general case the number of BSM  matrix elements for a given
chirality and isospin one has to evaluate at the factorization scale is reduced to 8. These are the matrix elements of
\be\label{8DQCD}
B_1,\qquad C_1, \qquad D_{1-3}, \qquad  D^*_{1-3}\,.
\ee
 However, in the chiral limit this number reduces to 3. Indeed, at the factorization scale the matrix elements of $D_{1-3}^*$ vanish and
we have the relations:
\be
\langle D_1(0)\rangle_I=\langle D_2(0)\rangle_I\,, \qquad \langle B_1(0)\rangle_I=-\langle D_3(0)\rangle_I\,,
\ee
such that it is sufficient to calculate
the  matrix elements of
\be\label{3DQCD}
B_1, \qquad  C_1, \qquad   D_1\,.
\ee

But above the factorization scale the situation changes  as can be seen from the meson evolution. In particular,  the relation
between matrix elements of $B_1$ and $D_3$ is violated while the one between
$D_1$ and $D_2$ is preserved. Most importantly the matrix elements of
tensor-tensor operators $D^*_{1-3}$ become non-zero.

\section{SD basis}\label{sec:4}
While the short distance evolution for scales above $\mu=1\gev$ can
be performed in the DQCD basis as demonstrated above,
for short distance renormalization group evolution a different basis of
13 BSM operators based on
 \cite{Buras:2000if} is more useful\footnote{We thank Mikolaj Misiak for discussions.}.
That basis, to be termed SD basis in what follows, is closer to
bases used in various computer codes present in the literature
(see Appendices~\ref{Flavio} and \ref{Flavio2}). So, for completeness we would like to present our results in that
basis as well. In the SD basis, like in the DQCD basis, no summations over quark flavours are performed, but in contrast to the DQCD basis,
colour non-singlet operators are present. While this is a disadvantage with respect to the DQCD basis as far as calculations of hadronic matrix elements are
concerned, it turns out to be more suitable for quark-gluon evolution.

The 13 BSM independent operators in the SD basis are also linearly independent from the SM ones, in particular none of them
mixes into QCD- and QED-penguin operators $Q_{3,\ldots 10}$.
Using the notation of  \cite{Buras:2000if}
they are given as follows:
\newline
\newline
{\bf Class I:}

\begin{equation}
\begin{aligned}
  Q_1^{{\rm VLL},d-s} & = (\bar s^\alpha \gamma_\mu P_L d^\alpha) \,
  \big[ (\bar d^\beta \gamma^\mu P_L d^\beta) - (\bar s^\beta \gamma^\mu P_L s^\beta)\big] ,
\end{aligned}
\end{equation}
\noindent
 {\bf Class II}

 \begin{align}
   Q_{1}^{{\rm SLR},u} &
   = (\bar s^\alpha P_L d^\beta) \, (\bar u^\beta P_R \, u^\alpha) ,\label{SLRu1}
 \\
   Q_{2}^{{\rm SLR},u} &
   = (\bar s^\alpha P_L d^\alpha) \, (\bar u^\beta P_R \, u^\beta) ,\label{SLRu2}
   \\
     Q_1^{{\rm VLR},d-s} & = (\bar s^\alpha \gamma_\mu P_L d^\beta) \,
     \big[ (\bar d^\beta \gamma^\mu P_R d^\alpha) - (\bar s^\beta \gamma^\mu P_R s^\alpha)\big] ,\label{VLR1}
   \\
     Q_2^{{\rm VLR},d-s} & = (\bar s^\alpha \gamma_\mu P_L d^\alpha) \,
     \big[ (\bar d^\beta \gamma^\mu P_R d^\beta) - (\bar s^\beta \gamma^\mu P_R s^\beta)\big] ,\label{VLR2}
 \end{align}
\newline
\newline
{\bf Class III}

\begin{align}
  Q_{1}^{{\rm SLL}, u} &
  = (\bar s^\alpha P_L d^\beta) \, (\bar u^\beta P_L \, u^\alpha) ,
\\
  Q_{2}^{{\rm SLL}, u} &
  = (\bar s^\alpha P_L d^\alpha) \, (\bar u^\beta P_L \, u^\beta) ,
\\
  Q_{3}^{{\rm SLL}, u} &
  = - (\bar s^\alpha \sigma_{\mu\nu} P_L d^\beta) \, (\bar u^\beta \sigma^{\mu\nu} P_L \, u^\alpha) ,
\\
  Q_{4}^{{\rm SLL}, u} &
  = - (\bar s^\alpha \sigma_{\mu\nu} P_L d^\alpha) \, (\bar u^\beta \sigma^{\mu\nu} P_L \, u^\beta) ,
\end{align}
\newline
\newline
{\bf Class IV}

\begin{align}
  Q_{1}^{{\rm SLL}, d} &
  = (\bar s^\alpha P_L d^\beta) \, (\bar d^\beta P_L \, d^\alpha) ,
\\
  Q_{2}^{{\rm SLL}, d} &
  = (\bar s^\alpha P_L d^\alpha) \, (\bar d^\beta P_L \, d^\beta) ,
\end{align}

\begin{align}
  Q_{1}^{{\rm SLL}, s} &
  = (\bar s^\alpha P_L d^\beta) \, (\bar s^\beta P_L \, s^\alpha) ,
\\
  Q_{2}^{{\rm SLL}, s} &
  = (\bar s^\alpha P_L d^\alpha) \, (\bar s^\beta P_L \, s^\beta) .
\end{align}

The connection between the operators in the DQCD and
SD bases is rather simple and  given in Appendix~\ref{App}. To obtain these
relations, Fierz identities in (\ref{F1})-(\ref{F5}) have been used. Having
these relations and the expressions for meson evolution in the DQCD
basis, it is straightforward to obtain analogous evolution in the SD
basis. It is given in Appendix~\ref{MEinSD}.
The large $N$ hadronic matrix elements of the 13 BSM operators in the SD basis are collected
in Appendix~\ref{sec:BBM} and one-loop anomalous dimension
matrices in Appendix~\ref{SDADM}.

Similar to the discussion in Section~\ref{Counting}, the number of matrix
elements one has to calculate at the factorization scale is reduced for
the following reasons. As
$\bar s s$ bilinears do not contribute to the matrix elements at the factorization scale, the matrix elements of $Q_{1,2}^{{\rm SLL}, s}$ vanish while those of  $Q_1^{{\rm VLL},d-s}$ and  $Q_{1,2}^{{\rm VLR},d-s}$ can be then expressed in terms of the SM ones as follows
\be
\langle Q_1^{{\rm VLL},d-s}\rangle_I= \frac{1}{6} (\langle Q_3\rangle_I-\langle Q_9\rangle_I),
\ee
\be
\langle Q_1^{{\rm VLR},d-s}\rangle_I= \frac{1}{6} (\langle Q_6\rangle_I-\langle Q_8\rangle_I),
\ee
\be
\langle Q_2^{{\rm VLR},d-s}\rangle_I= \frac{1}{6} (\langle Q_5\rangle_I-\langle Q_7\rangle_I)\,,
\ee
wherever $\langle \bar s \Gamma_A s\rangle=0$.
Therefore, in the general case the number of BSM  matrix elements for a given
chirality and isospin one has to evaluate at the factorization scale is reduced
as in the DQCD basis to 8:
\be\label{8SD}
 Q_{1,2}^{{\rm SLR},u}\,, \qquad  Q_{1-4}^{{\rm SLL},u}, \qquad  Q_{1,2}^{{\rm SLL},d}\,.
\ee

However in the chiral limit, analogous arguments to those presented in
 Section~\ref{Counting} imply that only the matrix elements of
the following operators have to be evaluated
\be\label{3SD}
 Q_{1}^{{\rm SLR},u}\,, \qquad  Q_{2}^{{\rm SLR},u}, \qquad Q_{1}^{{\rm SLL}, u}\,.
\ee
This can be easily verified by using the result in (\ref{3DQCD}) together
with the connection between the operators in the DQCD and
SD bases given in Appendix~\ref{App}.

\section{SMEFT view of BSM operators}\label{sec:SMEFT}
{Following \cite{Aebischer:2018quc,Aebischer:2018csl},
it should be emphasized that while generally  13  BSM operators
are consistent with the $\text{SU(3)}_c\times U(1)_{\text{Q}}$ symmetry, only
7 operators are consistent with the full SM gauge symmetry $\text{SU(3)}_c\times\text{SU(2)}_L\times U(1)_{\text{Y}}$ \cite{Aebischer:2015fzz,Jenkins:2017jig}\footnote{We thank Christoph Bobeth and David Straub for discussions.}. As in \cite{Aebischer:2018csl} a different operator basis has been used, we
identify here these 7 operators in the SD and DQCD bases.}

Beginning with the
SD basis, the operators in (\ref{SLRu1}) and (\ref{SLRu2}) and in class IV violate the full SM gauge symmetry $\text{SU(3)}_c\times\text{SU(2)}_L\times U(1)_{\text{Y}}$ \cite{Aebischer:2015fzz,Jenkins:2017jig}. In particular one can easily check that they
do not conserve hypercharge. Consequently, if no new particles close to
electroweak scale exist and the SMEFT is the correct
description between the electroweak scale and the NP scale, then
the Wilson coefficients of the operator (\ref{SLRu2}) and those in class IV must vanish. Though explicitly carrying a non-zero weak hypercharge ($Y=2$), the operator (\ref{SLRu1}) can in principle be generated through the dimension-six gauge-invariant operator $i (\tilde \phi^\dagger D_{\mu}\phi)(\bar u\gamma^{\mu}P_Rd)$ frozen at the vacuum expectation value of the Higgs doublet $\phi$ ($Y=1$). Such would for example be the case in an $SU(2)_L\times SU(2)_R\times U(1)_{B-L}$ UV completion of the SM via the tree-level $W_L-W_R$ mixing. If we neglect this modified $W_L$ coupling \cite{Aebischer:2015fzz},
 the full set of contributing linearly independent operators contains not 40 but 28 operators:
\begin{itemize}
\item
7 SM operators and the corresponding 7 mirror operators with L and R interchanged,
\item
7 BSM operators, those in classes I and III and the operators in (\ref{VLR1})
and (\ref{VLR2}), and the corresponding 7 mirror operators obtained again by interchanging L and R.
\end{itemize}
 Similarly, in the DQCD basis the imposition of invariance under the full gauge group of the SM reduces the number of contributing BSM operators of a given chirality to 7. In Table~\ref{tab:op-counting} we list these operators.

\begin{table}[t]
\centering
\renewcommand{\arraystretch}{1.3}
\begin{tabular}{lcccccc}
\toprule
          & VLL     & SLR,u     &  VLR  & SLL,u     & SLL,d     & SLL,s      \\
\hline
  SD
          & $Q_1^{{\rm VLL},d-s}$ & $--$ & $Q_{1,2}^{{\rm VLR},d-s}$ & $Q_{1-4}^{{\rm SLL}, u}$ & $--$ & $--$ \\
  DQCD   & A & $--$ & $B_2,C_2$ & $D^{(*)}_{1,2}$ & $--$ & $--$ \\
\bottomrule
\end{tabular}
\caption{SD and DQCD BSM operators generated in SMEFT.}
  \label{tab:op-counting}
\end{table}

As discussed already at the end of  Sections~\ref{Counting} and \ref{sec:4}, the number
of matrix elements one has to evaluate at the factorization scale for given chirality and isospin is reduced
to 8 when one takes into account that
 $\bar s s$ bilinears do not contribute to the matrix elements at this scale.
They are listed in (\ref{8DQCD}) and (\ref{8SD}) for the DQCD and SD bases,
respectively.  Evidently then in the SMEFT this number is reduced to 4:
\be
\text{DQCD}:~~D^{(*)}_{1,2},\qquad \text{SD}:~~Q_{1-4}^{{\rm SLL}, u}
\ee

However, in the chiral limit even without the imposition of SMEFT, the number
of matrix elements to be calculated at the factorization scale in each basis
is reduced to 3. They are given in (\ref{3DQCD}) and (\ref{3SD}). Consequently, once SMEFT is imposed only one matrix element at the factorization scale
in each basis for a given chirality and isospin has to be evaluated. For instance
\be
\text{DQCD}:~~D_1,\qquad \text{SD}:~~Q_{1}^{{\rm SLL}, u}\, \qquad (\text{SMEFT}).
\ee

Once meson evolution is turned on the picture is much reacher as some
relations between matrix elements valid at the factorization scale are broken.
Most important, the matrix elements of colour singlet tensor-tensor operators
 do not vanish any longer.

\section{Numerical Results}\label{sec:6}
Having all these results for hadronic matrix elements in DQCD and SD bases
at hand, we will next present their numerical values. To this end we will
assume that values of the matrix elements at $\Lambda=0.7\gev$ give an adequate
representation of their values at $1\gev$. This treatment was rather successful
in the case of our analysis of $K^0-\bar K^0$ matrix elements in \cite{Buras:2018lgu} and we expect that it is a reasonable approximation for the time
being. This is furthermore supported by the fact, as discussed at the end of  Section~\ref{sec:5},
that the meson evolutions for operators discussed in the present paper have
similar structure to the ones in \cite{Buras:2018lgu}.

Next two features should be noticed:
\begin{itemize}
\item The BSM hadronic matrix elements of classes B and D, involving the factor $r^2(\mu)$,
are chirally enhanced with
\be
r^2(1\gev)\approx 60\,(m_K^2-m_\pi^2)
\ee
and
 the corresponding operators have the highest potential to have an
impact on $\epe$ without requiring large values of their Wilson coefficients.
\item Among the chirally enhanced matrix elements of classes B and D, those contributing to the
isospin amplitude $A_2$ are most important as the contribution of the amplitude
$A_0$ to $\epe$ is automatically suppressed by a factor $1/22$ relative
to the one of $A_2$ since
\begin{align}
  \label{eq:epe-formula}
  \epe &
  \propto
    \left[\frac{\text{Re}A_2}{\text{Re}A_0} \text{Im}A_0
         - \text{Im}A_2 \right]\,, \qquad \frac{\text{Re}A_2}{\text{Re}A_0}\approx \frac{1}{22}.
\end{align}

Consequently
as pointed out in \cite{Buras:2015jaq},
in order to obtain a significant contribution to this ratio from NP contributing
to $A_0$ the imaginary part of the corresponding Wilson coefficient should be larger than in the case of $A_2$
 in order to compensate this suppression. This in turn could lead, in certain NP models, to the violation of the present bounds on rare decays.
\item
The values of  $I=0$ and $I=2$ matrix elements of all BSM operators are  at $\mu=\ord(1\gev)$ similar
to each other so that we do not expect these new operators to be relevant
for the $\Delta I=1/2$ rule. This is an important result as it implies that either this rule is fully governed by the SM dynamics or NP contributions to
the $A_0$ amplitude,  at the level of $(10-20)\%$,  come from modifications of
the Wilson coefficients of the SM operators. See \cite{Buras:2014sba} for a
detailed analysis.
\end{itemize}

Now among the hadronic matrix elements of SM penguin operators calculated by lattice QCD and DQCD, the most important ones are
the matrix elements of $Q_6$ and $Q_8$ which
are given as follows  \cite{Buras:1985yx,Bardeen:1986vp,Buras:1987wc}
\be\label{Q60}
\langle Q_6(\mu) \rangle_0 =-\, r^2(\mu)  (F_K-F_\pi)\bsi,\qquad \bsi=0.59\pm
0.19,
\ee
\be\label{Q82}
\langle Q_8(\mu) \rangle_2 =\frac{1}{2\sqrt{2}}
r^2(\mu) F_\pi \bei\,, \qquad \bei=0.76\pm 0.05\,,
\ee
with the values of $\bsi$ and $\bei$ from RBC-UKQCD collaboration and similar
results from DQCD \cite{Buras:2015xba}.
It will be then
of interest to compare the values of the matrix elements of BSM operators
with these two matrix elements for $A_0$ and $A_2$, respectively.
In doing this we have to take into account that $Q_6$ and $Q_8$
have the $(V-A)\otimes (V+A)$ Dirac structure, while the BSM operators considered by us involve  the $P_L$ and $P_R$ chiral projectors. In order to compensate for this, we will
give the results for
\be
\langle \hat Q_6(\mu)\rangle_0 \equiv \frac{\langle Q_6(\mu) \rangle_0}{4},\qquad
\langle \hat Q_8(\mu)\rangle_2 \equiv \frac{\langle Q_8(\mu) \rangle_2}{4}\,.
\ee

For the numerical analysis of the matrix elements SD evolution we use \texttt{WCxf} \cite{Aebischer:2017ugx} and \textsf{\emph{wilson}} \cite{Aebischer:2018bkb} as well as the results from the previous section. Below the hadronic scale, the running of the matrix elements is given by the
meson evolution formulae given in previous sections.

We use the following input
\be
F=130.41\mev,\quad m_K=497.6\mev,\quad m_\pi=134.98\mev,\qquad \alpha_s(M_Z)=0.1181\, .
\ee

In Tables~\ref{tab:SMEFT-DQCD} and \ref{tab:U(1)-DQCD} we show the results in
the DQCD basis for operators allowed by SMEFT and forbidden by it, respectively.
The corresponding results for the SD basis are given in
Tables~\ref{tab:SMEFT-SD} and \ref{tab:U(1)-SD}.
 Let us then extract the most
important lessons from  these tables.
\begin{itemize}
\item
Concentrating first on $I=2$ matrix elements, that according to (\ref{eq:epe-formula}) could have larger impact on $\epe$ than $I=0$ matrix elements,
we pin down several BSM operators for which the values of matrix elements are
in the ballpark of the value of
$\langle Q_8(\mu) \rangle_2$ for $\mu=\ord(1\gev)$. As the Wilson
coefficient of $Q_8$ is $\ord(\alpha_{\text{e}})$, it is conceivable that some
of these operators could have significant impact on $\epe$.

In the DQCD basis this is the case of the pure tensor-tensor operators
\be\label{Dwinners}
D_1^*,\qquad D_2^*, \qquad D_3^*{\,,} \qquad (\text{DQCD~basis})
\ee
with  the first two allowed by SMEFT.

In the SD basis this is the case of the operators
\be
Q_1^{{\rm VLR},d-s}, \qquad  Q_{3}^{{\rm SLL}, u},\qquad  Q_{4}^{{\rm SLL}, u}{\,,} \qquad  (\text{SD~basis})
\ee
with all three allowed by SMEFT.
\item
Looking then at $I=0$ matrix elements, we bring out a number of matrix elements
which are comparable to or larger than the one of $\langle Q_6(\mu) \rangle_0$. In the DQCD
basis this is the case not only for the operators in (\ref{Dwinners}) but
also for $B_{1,2}$ and $D_3$. In the case of the SD basis the additional large $I=0$
matrix elements are those of  $ Q_{2}^{{\rm SLR},u}$ and  $ Q_{2}^{{\rm SLL},d}$.
\end{itemize}

Explicitly we have
\be\label{N11}
|\langle D^*_{1,2} \rangle_2|\approx \langle Q_{4}^{{\rm SLL}, u}\rangle_2 \approx
1.2~ |\langle Q_8(\mu) \rangle_2|,
\ee
\be\label{N22}
|\langle D^*_{3} \rangle_2|\approx
0.8~ |\langle Q_8(\mu) \rangle_2|,\qquad
|\langle Q_{3}^{{\rm SLL}, u}\rangle_2|\approx
2.0~ |\langle Q_8(\mu) \rangle_2|\,,
\ee
at $\mu=1\gev$.
In the DQCD basis the meson evolution from the factorization scale to scales $\ord(1\gev)$ is primarily responsible for this result. It should however be
noticed that while the matrix elements of $Q_6$, $Q_8$ and generally scalar-scalar operators increase with increasing $\mu$, the ones of
tensor-tensor operators are only weakly dependent on $\mu$ for $\mu > 1\gev$.
Therefore the numerical factors in (\ref{N11}) and (\ref{N22}) generally decrease significantly with increasing $\mu$. An exception is the operator $ Q_{3}^{{\rm SLL}, u}$  which, among tensor-tensor operators in both operator bases, is the only one which is colour-non-singlet. We will return to this point at the end of the present section.

It is of interest to understand why the matrix elements of the remaining tensor-tensor operators exhibit such a weak dependence on $\mu$. This is most transparently seen
by studying the RG formula (\ref{RGOi}) for $X_i=D^*_1$ together with the ADM
in (\ref{eq:CIII}), although it can already be suspected at the one-loop level, from (\ref{eq:Dst12}) and (\ref{D*D12}) which imply $\langle D_{2,1}-\frac{1}{4}D_{2,1}^*\rangle=0$ at $\Lambda=2\pi F \approx 0.8$ GeV.

Now the matrix element $\langle D_1^*(\mu_1)\rangle$ with $\mu_1=1\gev$  is
generated, as seen in (\ref{D*D12}),  in the process of meson
evolution, through mixing
 with the scalar-scalar operator $D_2$. Because this mixing
is very large and further enhanced through short but fast meson evolution
$\langle D_1^*(\mu_1)\rangle$ is, as seen in Table~\ref{tab:SMEFT-DQCD}, much larger than $\langle D_2(\mu_1)\rangle$. The values of $\langle D_1^*(\mu_2)\rangle$,
for $\mu_2>\mu_1$ are now governed first of all by the self-mixing of $D^*_1$,
the $(3,3)$ in (\ref{eq:CIII}) and the mixing of $D_1^*$ with $D_2$ given by
the entry $(3,2)$ in (\ref{eq:CIII}). But while the $(3,3)$ entry is much smaller
than $(3,2)$, the matrix element of  $D_1^*$ is much larger than that of $D_2$
as mentioned above. Using this information in (\ref{RGOi}) we find that these
effects cancel each other to a large extend leaving a very weak $\mu$ dependence
of $D_1^*(\mu)$.  This cancellation even improves with increasing $\mu$ because,
while the diagonal evolution of $D_1^*$ slowly decreases its matrix element,
the one of $D_2$ governed by the large entry $(2,2)$ in (\ref{eq:CIII}) and
having opposite sign to $(3,3)$ increases this matrix element. The evolution of $D^*_1$
is governed in addition by the entry $(3,1)$, the mixing of $D_1^*$ and $D_1$,
and by the entry $(3,4)$, the mixing of $D_1^*$ and $D^*_2$. Taking into account that the matrix elements of $D_2^*$ and  $D_1^*$ are equal to each other and the same applies to  matrix elements of $D_1$ and  $D_2$, one can then easily check using  (\ref{RGOi}) that these two additional effects cancel each other to a large extent and have only a very small impact on the evolution of $D^*_1$. This result
remains true after performing the summation of leading logarithms to all orders
of perturbation theory,  which all the numerical values in the tables are based on.

\begin{table}
\centering
\captionsetup{width=0.89\textwidth}
\begin{tabular}{||c|c|c|c|c|c|c|c|c||}
\hline
 $I=0$
& $A$ & $B_2$ & $C_2$ & $D_1$ & $D_2$ & $D^*_1$ & $D^*_2$& $\hat Q_6$\\
\hline
$1\gev$ & 0.001 & 0.058 & -0.006 & -0.039 & -0.039 & -0.224 & -0.224 & -0.053 \\
$1.3\gev$ & 0.001 & 0.070 & -0.015 &-0.044 & -0.044 & -0.213 & -0.214 & -0.071 \\
$2\gev$  & 0.001 & 0.088 & -0.028 & -0.050 & -0.050 & -0.207 & -0.207 & -0.099 \\
$3\gev$  & 0.001 & 0.104 & -0.039 & -0.055 & -0.055 & -0.204 & -0.204 & -0.125 \\
\hline\hline
$I=2$
& $A$ & $B_2$ & $C_2$ & $D_1$ & $D_2$ & $D^*_1$ & $D^*_2$& $\hat Q_8$\\
\hline
$1\gev$  & -0.001 & 0.041 & 0.001 & -0.028 & -0.028 & -0.158 & -0.158 & 0.124 \\
$1.3\gev$ & -0.001 & 0.050 & -0.006 & -0.031 & -0.031 & -0.151 & -0.151 & 0.164 \\
$2\gev$  & -0.001 & 0.062 & -0.015 & -0.035 & -0.035 & -0.146 & -0.146 & 0.230 \\
$3\gev$  & -0.001 & 0.074 & -0.023 & -0.039 & -0.039 & -0.145 & -0.145 & 0.290 \\
\hline
\end{tabular}
\renewcommand{\arraystretch}{1.3}
\caption{\label{tab:SMEFT-DQCD}
Matrix elements $\langle X_i\rangle_{0,2}$ of BSM operators in the DQCD basis allowed by SMEFT contributing to the  isospin amplitudes $A_{0,2}$ in units of $\gev^3$ for four values of $\mu$. In the last column we give the values of the matrix
elements $\langle \hat Q_6(\mu)\rangle_0$ and $\langle \hat Q_8(\mu)\rangle_2$
for comparison.
}
\end{table}

\begin{table}
\centering
\captionsetup{width=0.89\textwidth}
\begin{tabular}{||c|c|c|c|c|c|c|c||}
\hline
 $I=0$
& $B_1$ & $C_1$ &  $D_3$ & $D_4$ & $D^*_3$ & $D^*_4$ & $\hat Q_6$ \\
\hline
$1\gev$  & 0.116 & -0.012 & -0.079 & 0 & -0.298 & 0 & -0.053\\
$1.3\gev$ & 0.141 & -0.030 & -0.088 & 0  & -0.298 & 0 & -0.071\\
$2\gev$  & 0.176 & -0.056 & -0.101 & 0 & -0.303 & 0 & -0.099\\
$3\gev$  & 0.208 & -0.078 & -0.111 & 0 & -0.311 & 0 & -0.125\\
\hline\hline
$I=2$
& $B_1$ & $C_1$ &  $D_3$ & $D_4$ & $D^*_3$ & $D^*_4$ & $\hat Q_8$\\
\hline
$1\gev$  & -0.041 & -0.001 & 0.028 & 0 & 0.105 & 0 & 0.124 \\
$1.3\gev$ & -0.050 & 0.006 & 0.031 & 0 & 0.105 & 0 & 0.164 \\
$2\gev$  & -0.062 & 0.015 & 0.036 & 0 & 0.107 & 0 & 0.230 \\
$3\gev$  & -0.074 & 0.023 & 0.039 & 0 & 0.110 & 0 & 0.290 \\
\hline
\end{tabular}
\renewcommand{\arraystretch}{1.3}
\caption{\label{tab:U(1)-DQCD}
Matrix elements $\langle X_i\rangle_{0,2}$ of BSM operators in the DQCD basis allowed by $\text{SU(3)}_c\times U(1)_{\text{Q}}$  but not by SMEFT contributing to the  isospin amplitudes $A_{0,2}$ in units of $\gev^3$ for four values of $\mu$.  In the last column we give the values of the matrix
elements $\langle \hat Q_6(\mu)\rangle_0$ and $\langle \hat Q_8(\mu)\rangle_2$
for comparison.
}
\end{table}

\begin{table}
\centering
\captionsetup{width=0.89\textwidth}
\begin{tabular}{||c|c|c|c|c|c|c|c||}
\hline
 $I=0$
& $ Q_1^{{\rm VLL},d-s}$ & $ Q_1^{{\rm VLR},d-s}$ & $ Q_2^{{\rm VLR},d-s}$ & $ Q_{1}^{{\rm SLL}, u}$ & $ Q_{2}^{{\rm SLL}, u}$ & $ Q_{3}^{{\rm SLL}, u}$ & $ Q_{4}^{{\rm SLL}, u}$\\
\hline
$1\gev$ & 0.001 & -0.116 & -0.006 & -0.008 & -0.039 & -0.348 & -0.224 \\
$1.3\gev$ & 0.001 & -0.141 & -0.015 & -0.005 & -0.044 & -0.371 & -0.214 \\
$2\gev$ & 0.001 & -0.176 & -0.028 & -0.001 & -0.050 & -0.404 & -0.207 \\
$3\gev$ & 0.001 & -0.208 & -0.039 & 0.002 & -0.055 & -0.433 & -0.204 \\
\hline\hline
$I=2$
& $ Q_1^{{\rm VLL},d-s}$ & $ Q_1^{{\rm VLR},d-s}$ & $ Q_2^{{\rm VLR},d-s}$ & $ Q_{1}^{{\rm SLL}, u}$ & $ Q_{2}^{{\rm SLL}, u}$ & $ Q_{3}^{{\rm SLL}, u}$ & $ Q_{4}^{{\rm SLL}, u}$\\
\hline
$1\gev$ & -0.001 & -0.082 & 0.001 & -0.006 & -0.028 & -0.246 & -0.158 \\
$1.3\gev$ & -0.001 & -0.100 & -0.006 & -0.003 & -0.031 & -0.262 & -0.151 \\
$2\gev$ & -0.001 & -0.125 & -0.015 & -0.001 & -0.035 & -0.285 & -0.146 \\
$3\gev$ & -0.001 & -0.147 & -0.023 & 0.001 & -0.039 & -0.306 & -0.145 \\
\hline
\end{tabular}
\renewcommand{\arraystretch}{1.3}
\caption{\label{tab:SMEFT-SD}
Matrix elements $\langle X_i\rangle_{0,2}$ of BSM operators in the SD basis allowed by SMEFT contributing to the  isospin amplitudes $A_{0,2}$ in units of $\gev^3$ for four values of $\mu$.
}
\end{table}

\begin{table}
\centering
\captionsetup{width=0.89\textwidth}
\begin{tabular}{||c|c|c|c|c|c|c||}
\hline
 $I=0$
& $ Q_{1}^{{\rm SLR},u}$ & $ Q_{2}^{{\rm SLR},u}$ &  $ Q_{1}^{{\rm SLL}, d}$ & $ Q_{2}^{{\rm SLL}, d}$ & $ Q_{1}^{{\rm SLL}, s}$ & $ Q_{2}^{{\rm SLL}, s}$\\
\hline
$1\gev$ & -0.006 & -0.116 & 0.002 & -0.079 & 0 & 0 \\
$1.3\gev$ & -0.015 & -0.141 & 0.007 & -0.088 & 0 & 0 \\
$2\gev$ & -0.028 & -0.176 & 0.012 & -0.101 & 0 & 0 \\
$3\gev$ & -0.039 & -0.208 & 0.017 & -0.111 & 0 & 0 \\
\hline\hline
$I=2$
& $ Q_{1}^{{\rm SLR},u}$ & $ Q_{2}^{{\rm SLR},u}$ &  $ Q_{1}^{{\rm SLL}, d}$ & $ Q_{2}^{{\rm SLL}, d}$ & $ Q_{1}^{{\rm SLL}, s}$ & $ Q_{2}^{{\rm SLL}, s}$\\
\hline
$1\gev$ & 0.000 & 0.041 & -0.001 & 0.028 & 0 & 0 \\
$1.3\gev$ & 0.003 & 0.050 & -0.002 & 0.031 & 0 & 0 \\
$2\gev$ & 0.007 & 0.062 & -0.004 & 0.036 & 0 & 0 \\
$3\gev$ & 0.011 & 0.074 & -0.006 & 0.039 & 0 & 0 \\
\hline
\end{tabular}
\renewcommand{\arraystretch}{1.3}
\caption{\label{tab:U(1)-SD}
Matrix elements $\langle X_i\rangle_{0,2}$ of BSM operators in the SD basis allowed by $\text{SU(3)}_c\times U(1)_{\text{Q}}$  but not by SMEFT contributing to the  isospin amplitudes $A_{0,2}$ in units of $\gev^3$ for four values of $\mu$.
}
\end{table}

In order to further demonstrate that the pattern of meson evolution
agrees with the SD evolution, it is
useful
to normalize  our results for BSM matrix elements of scalar-scalar operators to $\langle \hat Q_8(\mu) \rangle_2$ and consider the ratios
\be\label{Riratios}
R_I(X_i(\mu))= \,\frac{\langle X_i(\mu)\rangle_I}{\langle \hat Q_8(\mu) \rangle_2}\,
\qquad (\text{scalar-scalar})
\ee
with $X_i$ denoting any scalar-scalar operator either in DQCD or SD basis.
This has the virtue that in the case of chirally enhanced matrix elements
of scalar-scalar operators the
dominant $\mu$-dependence present in $r(\mu)$ cancels out, exhibiting
the  non-factorizable $\mu$ dependence of $R_I$ for such operators.

Inspecting the Tables~\ref{tab:RIDQCDa} and \ref{tab:RIDQCDb} for the DQCD
 basis and  Tables~\ref{tab:SMEFT-SDa} and \ref{tab:SMEFT-SDb}
for the SD basis,  we can make the following observations:
\begin{itemize}
\item
The ratios $R_I$ for  scalar-scalar matrix elements governed by $r^2(\mu)$,
that is of $B_{1,2}$ and $D_{1,2,3}$ in the DQCD basis, all decrease with increasing $\mu$ so that
the SD evolution matches well the meson one. In particular, the SD
evolution of $B_{1,2}$ is in the range $1-3\gev$ by roughly a factor of two
slower than that of $D_{1,2,3}$ in agreement with the meson evolution equations of Section~\ref{sec:5}. The matrix elements of $A$ and $C_{1,2}$ are not chirally
enhanced and the ratios $R_I$ are not useful in this case for exhibiting
the proper matching of SD and meson evolutions. However, inspecting the SD and
meson evolutions, also in this case the matching is good.
\item
The ratios $R_I$ for scalar-scalar matrix elements of $Q_1^{{\rm VLR},d-s}$,
$ Q_{1,2}^{{\rm SLL}, u}$,  $ Q_{2}^{{\rm SLR},u}$, $ Q_{1,2}^{{\rm SLL}, d}$ in the
SD basis, all governed by $r^2(\mu)$, exhibit also the SD pattern as expected
from the matching of SD evolution to the meson evolution summarized in
Appendix~\ref{MEinSD}. All decrease with increasing $\mu$ with the speed
expected from meson evolution. The comments made on matrix elements not enhanced by $r^2(\mu)$ in the DQCD basis applies also to the SD basis.
\end{itemize}

In the case of tensor-tensor operators the diagonal evolution of these operators
differs from the one of scalar-scalar operators simply because their
anomalous dimensions, as seen from (\ref{AST}), are very different and read
for $N=3$
\be
\gamma(Q_8)=2\gamma_S=-16,\qquad \gamma(Q_T)=2\gamma_T=\frac{16}{3},
\ee
where $Q_T$ denotes any colour singlet tensor-tensor operator.
Consequently, the diagonal SD evolutions of $\langle Q_8(\mu) \rangle$ and
$\langle Q_T(\mu) \rangle$ are rather different
\be
\langle Q_8(\mu) \rangle \propto [\alpha_s(\mu)]^{\gamma_S/\beta_0},\qquad
\langle Q_T(\mu)\rangle \propto [\alpha_s(\mu)]^{\gamma_T/\beta_0}\,,
\ee
with $\beta_0=11-2f/3$. Moreover, as we discussed above, the evolution of
tensor-tensor operators is not governed by their diagonal evolution but rather
by a complicated ADM. Therefore,  the ratios in (\ref{Riratios})  are not useful in this case for
the demonstration of the matching of non-factorizable evolutions and  we do not show them in  Tables \ref{tab:RIDQCDa}-\ref{tab:SMEFT-SDa}.

An exception is the colour non-singlet operator $Q_{3}^{{\rm SLL}, u}$, as already
mentioned above. In this case, in order to exhibit better non-factorizable SD evolution of matrix elements of this operator,  it appears to be more appropriate to consider the ratio
\be\label{Riratios2}
R_I(Q_{3}^{{\rm SLL}, u})= \frac{\langle Q_{3}^{{\rm SLL}, u}(\mu)\rangle_I}{\langle \hat Q_8(\mu) \rangle_2}\,
[\alpha_s(\mu)]^{(2\gamma_S-\gamma_3)/2\beta_0},
\qquad \gamma_3\equiv \gamma(Q_{3}^{{\rm SLL}, u})=-\frac{38}{3}
\ee
with $\gamma_3$  extracted from the $(3,3)$ entry in (\ref{ga0SLL}).
We show this ratio in Table~\ref{tab:SMEFT-SDa}. One can easily check that
the decrease of this ratio with increasing $\mu$ matches well the meson evolution of this operator. It is also interesting to note that at the one-loop level (\ref{eq:Dst34}) together with (\ref{D*D34}) imply $\langle D_3-D_3^*/4\rangle=0$ at a higher value of $\Lambda=\sqrt{6}\pi F= 1$ GeV compared to
 (\ref{eq:Dst12}) and (\ref{D*D12}) for which a similar cancellation occurs at
$\Lambda=2\pi F\approx 0.8$ GeV.

\begin{table}
\centering
\captionsetup{width=0.89\textwidth}
\begin{tabular}{||c|c|c|c|c|c||}
\hline
 $R_0(X_i(\mu))$
& $A$ & $B_2$ & $C_2$ & $D_1$ & $D_2$ \\
\hline
$0\gev$ & 0.020 & 0.620 & -0.020 & -0.620 & -0.620  \\
$1\gev$ & 0.005 & 0.469 & -0.050 & -0.318 & -0.318  \\
$1.3\gev$ & 0.004 & 0.429 & -0.093 & -0.268 & -0.268  \\
$2\gev$ & 0.003 & 0.384 & -0.121 & -0.218 & -0.218  \\
$3\gev$ & 0.002 & 0.359 & -0.135 & -0.190 & -0.190  \\
\hline\hline
 $R_2(X_i(\mu))$
& $A$ & $B_2$ & $C_2$ & $D_1$ & $D_2$ \\
\hline
$0\gev$ & -0.029 & 0.439 & 0.029 & -0.439 & -0.439  \\
$1\gev$ & -0.008 & 0.332 & 0.008 & -0.225 & -0.225 \\
$1.3\gev$ & -0.006 & 0.303 & -0.034 & -0.189 & -0.189  \\
$2\gev$ & -0.004 & 0.271 & -0.064 & -0.154 & -0.154  \\
$3\gev$ & -0.003 & 0.254 & -0.079 & -0.134 & -0.134 \\
\hline
\end{tabular}
\renewcommand{\arraystretch}{1.3}
\caption{\label{tab:RIDQCDa}
{Ratios $R_I$ of BSM operators in the DQCD basis allowed by SMEFT over $\hat Q_8$ contributing to the isospin amplitudes $A_{0,2}$ for four values of $\mu$.}}
\end{table}

\begin{table}
\centering
\captionsetup{width=0.89\textwidth}
\begin{tabular}{||c|c|c|c|c||}
\hline
 $R_0(X_i(\mu))$
& $B_1$ & $C_1$ &  $D_3$ & $D_4$ \\
\hline
$0\gev$ & 1.241 & -0.040 & -1.241 & 0 \\
$1\gev$ & 0.939 & -0.099 & -0.637 & 0  \\
$1.3\gev$ & 0.858 & -0.185 & -0.537 & 0 \\
$2\gev$ & 0.767 & -0.243 & -0.438 & 0  \\
$3\gev$ & 0.718 & -0.270 & -0.383 & 0  \\
\hline\hline
 $R_2(X_i(\mu))$
& $B_1$ & $C_1$ &  $D_3$ & $D_4$ \\
\hline
$0\gev$ & -0.439 & -0.029 & 0.439 & 0 \\
$1\gev$ & -0.332 & -0.008 & 0.225 & 0  \\
$1.3\gev$ & -0.303 & 0.034 & 0.190 & 0  \\
$2\gev$ & -0.271 & 0.064 & 0.155 & 0  \\
$3\gev$ & -0.254 & 0.079 & 0.136 & 0 \\
\hline
\end{tabular}
\renewcommand{\arraystretch}{1.3}
\caption{\label{tab:RIDQCDb}
{Ratios $R_I$ of BSM operators in the DQCD basis allowed by $\text{SU(3)}_c\times U(1)_{\text{Q}}$  but not by SMEFT over $\hat Q_8$ contributing to the isospin amplitudes $A_{0,2}$ for four values of $\mu$.}
}
\end{table}

\begin{table}
\centering
\captionsetup{width=0.89\textwidth}
\begin{tabular}{||c|c|c|c|c||}
\hline
 $R_0(X_i(\mu))$
& $ Q_1^{{\rm VLL},d-s}$ & $ Q_1^{{\rm VLR},d-s}$ & $ Q_2^{{\rm VLR},d-s}$ & $ Q_{3}^{{\rm SLL}, u}$\\
\hline
$0\gev$ & 0.020 & -1.241 & -0.020 & -4.273 \\
$1\gev$ & 0.005 & -0.939 & -0.050 & -3.233 \\
$1.3\gev$ & 0.004 & -0.858 & -0.093 &  -2.736 \\
$2\gev$ & 0.003 & -0.767 & -0.121 &  -2.232 \\
$3\gev$ & 0.002 & -0.718 & -0.135 &  -1.965  \\
\hline\hline
 $R_2(X_i(\mu))$
& $ Q_1^{{\rm VLL},d-s}$ & $ Q_1^{{\rm VLR},d-s}$ & $ Q_2^{{\rm VLR},d-s}$ & $ Q_{3}^{{\rm SLL}, u}$ \\
\hline
$0\gev$ & -0.029 & -0.877 & 0.029 & -3.021 \\
$1\gev$ & -0.008 & -0.664 & 0.008 & -2.286  \\
$1.3\gev$ & -0.006 & -0.606 & -0.034 & -1.935  \\
$2\gev$ & -0.004 & -0.542 & -0.064 & -1.578 \\
$3\gev$ & -0.003 & -0.508 & -0.079 & -1.389  \\
\hline
\end{tabular}
\renewcommand{\arraystretch}{1.3}
\caption{\label{tab:SMEFT-SDa}
{Ratios $R_I$ of BSM operators in the SD basis allowed by SMEFT over $\hat Q_8$ contributing to the isospin amplitudes $A_{0,2}$ for four values of $\mu$.}
}
\end{table}

\begin{table}
\centering
\captionsetup{width=0.89\textwidth}
\begin{tabular}{||c|c|c|c|c|c|c||}
\hline
 $R_0(X_i(\mu))$
& $ Q_{1}^{{\rm SLR},u}$ & $ Q_{2}^{{\rm SLR},u}$ &  $Q_{1}^{{\rm SLL}, d}$ & $ Q_{2}^{{\rm SLL}, d}$ & $ Q_{1}^{{\rm SLL}, s}$ & $ Q_{2}^{{\rm SLL}, s}$\\
\hline
$0\gev$ & -0.020 & -1.241 & 0.620 & -1.241 & 0 & 0  \\
$1\gev$ & -0.050 & -0.939 & 0.017 & -0.637 & 0 & 0  \\
$1.3\gev$ & -0.093 & -0.858 & 0.042 & -0.537 & 0 & 0  \\
$2\gev$ & -0.121 & -0.767 & 0.054 & -0.438 & 0 & 0  \\
$3\gev$ & -0.135 & -0.718 & 0.058 & -0.383 & 0 & 0  \\
\hline\hline
 $R_2(X_i(\mu))$
& $ Q_{1}^{{\rm SLR},u}$ & $ Q_{2}^{{\rm SLR},u}$ &  $ Q_{1}^{{\rm SLL}, d}$ & $ Q_{2}^{{\rm SLL}, d}$ & $ Q_{1}^{{\rm SLL}, s}$ & $ Q_{2}^{{\rm SLL}, s}$\\
\hline
$0\gev$ & -0.014 & 0.439 & -0.219 & 0.439 & 0 & 0 \\
$1\gev$ & -0.004 & 0.332 & -0.006 & 0.225 & 0 & 0 \\
$1.3\gev$ & 0.017 & 0.303 & -0.015 & 0.190 & 0 & 0 \\
$2\gev$ & 0.032 & 0.271 & -0.019 & 0.155 & 0 & 0 \\
$3\gev$ & 0.039 & 0.254 & -0.020 & 0.136 & 0 & 0 \\
\hline
\end{tabular}
\renewcommand{\arraystretch}{1.3}
\caption{\label{tab:SMEFT-SDb}
{ Ratios $R_I$ of BSM operators in the SD basis allowed by $\text{SU(3)}_c\times U(1)_{\text{Q}}$  but not by SMEFT over $\hat Q_8$ contributing to the isospin amplitudes $A_{0,2}$ for four values of $\mu$.}
}
\end{table}

\section{Summary and outlook}\label{sec:7}
Motivated by the hints for NP contributing to the ratio $\epe$ we
have calculated hadronic matrix elements of 13 BSM four-quark operators in DQCD, including
the meson evolution in the chiral limit. This is the first calculation
of these matrix elements to date, thereby opening the road to the investigations
of $\epe$ and $K\to\pi\pi$ decays beyond  all BSM analyses found in
recent literature \cite{Buras:2014sba,Buras:2015yca,Blanke:2015wba,Buras:2015kwd,Buras:2016dxz,Buras:2015jaq,Kitahara:2016otd,Endo:2016aws,Endo:2016tnu,
Cirigliano:2016yhc,Alioli:2017ces,Bobeth:2016llm,Bobeth:2017xry,Crivellin:2017gks,Bobeth:2017ecx,Endo:2017ums,Haba:2018byj,Chen:2018ytc,Chen:2018vog,Matsuzaki:2018jui,Haba:2018rzf} in which
 NP  affected only Wilson coefficients of SM operators.

Our main messages to take home are the following ones:
\begin{itemize}
\item
The pattern of long
distance evolution (meson evolution) matches once more the one of short distance evolution (quark-gluon evolution), a property which to our knowledge cannot be presently achieved in any other analytical framework. It should be emphasized that this
important result has been obtained for 13 operators without any free parameter
 except possibly the physical cut-off $\Lambda$ which in any case has to be chosen in our framework in the ballpark of $0.7\gev$. See Section~\ref{sec:6}
 for details.
\item
Several matrix elements and in particular those
of tensor-tensor operators have values at $\mu=\ord(1\gev)$ in the ballpark
of the ones of the dominant electroweak penguin matrix element
$\langle Q_8(\mu) \rangle_2$. Therefore they could have large impact on
$\epe$.
\item
The mixing of the scalar-scalar operators into tensor-tensor operators
in the process of meson evolution is responsible for this result so that
the role of scalar-scalar operators should not be underestimated. They
can be most easily generated through tree-level exchanges of heavy colourless
and coloured scalars.
\end{itemize}

Clearly, in order to identify the most important operators, also their Wilson
coefficients in a given NP scenario must be known. But having to disposal large
$K\to\pi\pi$ matrix elements identified in our paper will, in some NP scenarios, facilitate the removal of $\epe$ anomaly without violating other constraints.
 A
detailed analysis of $\epe$ with and without the imposition of SMEFT and
concentrating rather on the structure of Wilson coefficients of
BSM operators resulting from the renormalization group effects is presented
in an accompanying paper \cite{Aebischer:2018csl} while a master formula
for $\epe$ beyond the SM has been recently presented in  \cite{Aebischer:2018quc}. Furthermore the hadronic matrix elements computed in this paper are implemented in the open-source codes flavio \cite{Straub:2018kue} and smelli \cite{Aebischer:2018iyb}, which allow for a numerical analysis including $\epe$.

Our calculation of meson evolution has been performed in the chiral limit
but, as we already mentioned in the Introduction, DQCD could reproduce some
lattice results in this approximation. Still it is desirable to extend
our work beyond chiral limit, a goal which could be reached before
lattice QCD calculations for the hadronic matrix elements in question
 are expected to be available.

In the case of confirmation of $\epe$ anomaly predicted in DQCD by more
precise  lattice QCD calculations, our results will play an important role in selecting
those NP models that can provide sufficient upward shift in $\epe$ in order
to explain the data.

On the other hand, if future lattice QCD calculations within the SM will confirm
the data on $\epe$, our results will put strong constraints on the parameters of a multitude of NP models.

\section*{Acknowledgements}
We would like to thank Christoph Bobeth and David Straub for discussions and
Christoph Bobeth for checking some of our
short distance calculations.
This research was done and financially  supported by the DFG cluster
of excellence ``Origin and Structure of the Universe''.

\appendix

\section{Basis transformations}\label{App}
\subsection{From SD to DQCD basis}
Here we list the basis transformation between the SD and  DQCD bases.
\subsubsection*{Class A}
\begin{equation}
A=Q_1^{{\rm VLL},d-s}.
\end{equation}
\subsubsection*{Class B}
\begin{align}
B_1=&~Q_2^{{\rm SRL},u}\,, & B_2=&-\frac{1}{2}Q_1^{{\rm VLR},d-s}\,.
\end{align}

\subsubsection*{Class C}
\begin{align}
C_1&= -2Q_1^{{\rm SRL},u}\,, & C_2&= Q_2^{{\rm VLR},d-s}\,.
\end{align}
\subsubsection*{Class D}
\begin{align}
D_1&= -\f{1}{2}Q_1^{{\rm SLL},u}+\f{1}{8}Q_3^{{\rm SLL},u}, & D_2&= Q_2^{{\rm SLL},u}\,,\\
D_3&= Q_2^{{\rm SLL},d}\,,& D_4&= Q_2^{{\rm SLL},s}\,,\\
D_1^*&= 6Q_1^{{\rm SLL},u}+\f{1}{2}Q_3^{{\rm SLL},u}\,,& D_2^*&=\,Q_4^{{\rm SLL},u},\\
D_3^*&= 8Q_1^{{\rm SLL},d}+4Q_2^{{\rm SLL},d}\,,& D_4^*&= 8Q_1^{{\rm SLL},s}+4Q_2^{{\rm SLL},s}\,.
\end{align}

\subsection{From DQCD to SD basis}\label{DQCDSD}

{\bf Class I}
\begin{equation}
  Q_1^{{\rm VLL},d-s}  =  A\,.
\end{equation}
{\bf Class II}
\begin{align}
Q_{1}^{{\rm SLR},u} &
=  -\frac{1}{2}C_1'\,,&
Q_{2}^{{\rm SLR},u} &
= B_1'\,,\\
Q_1^{{\rm VLR},d-s} & =  -2B_2\,,&
Q_2^{{\rm VLR},d-s} & = C_2\,,
\end{align}
{\bf Class III}
\begin{align}
Q_{1}^{{\rm SLL}, u} &
= -\frac{1}{2}D_1+\frac{1}{8}D_1^*\, ,&
Q_{2}^{{\rm SLL}, u} &
= D_2\,, \label{QD1}
\\
Q_{3}^{{\rm SLL}, u} &
= 6D_1+\frac{1}{2}D_1^*\, ,&
Q_{4}^{{\rm SLL}, u} &
= D_2^*\,, \label{QD2}
\end{align}
{\bf Class IV}
\begin{align}
  Q_{1}^{{\rm SLL}, d} &
  = -\frac{1}{2}D_3+\frac{1}{8}D_3^*\, ,&
  Q_{2}^{{\rm SLL}, d} &
  = D_3\, ,
    \\
      Q_{1}^{{\rm SLL}, s} &
      = -\frac{1}{2}D_4+\frac{1}{8}D_4^*\,.&
        Q_{2}^{{\rm SLL}, s} &
        = D_4\,,
\end{align}
with ' on $C_1,B_1$ meaning mirror partner ($L\leftrightarrow R$) and $^*$ on $D_{1,2,3,4}$ meaning tensor-tensor partner of $P_L-P_L$.
\subsection{From Flavio  to SD basis}\label{Flavio}
To use \textsf{\emph{wilson}} for the SD matrix element evolution we need to translate the used SM and BSM operators into a predefined basis. We choose the flavio basis \cite{Straub:2018kue}, where the operators are defined in \cite{WET-3}. The transformation from the flavio basis into SM basis reads:
\begin{align}
Q_{1}^{\dag} &=4 O_{sduu}^{V,LL},
&
Q_{2}^{\dag} &=4\widetilde O_{sduu}^{V,LL},
\\
Q_{3}^{\dag} &=4 O_{sduu}^{V,LL}+4 O_{sddd}^{V,LL}+4 O_{sdss}^{V,LL},
&
Q_{5}^{\dag} &=4 O_{sduu}^{V,LR}+4 O_{sddd}^{V,LR}+4 O_{sdss}^{V,LR},
\\
Q_{6}^{\dag} &=4\widetilde O_{sduu}^{V,LR}-8 O_{sddd}^{S,{LR}}-8 O_{sdss}^{S,RL},
&
Q_{7}^{\dag} &=4O_{sduu}^{V,LR}-2 O_{sddd}^{V,LR}-2 O_{sdss}^{V,LR},
\\
Q_{8}^{\dag} &=4\widetilde O_{sduu}^{V,LR}+4 O_{sddd}^{S,{LR}}+4 O_{sdss}^{S,RL},
\end{align}
with $\tilde O_i$ meaning coloured partners of $O_i$.

Similarly, for the BSM operators one finds:
\\[1cm]
{\bf Class I}
\begin{equation}
  (Q_1^{{\rm VLL},d-s})^{\dag}  = O_{sddd}^{V,LL}-O_{sdss}^{V,LL} \,,
\end{equation}
{\bf Class II}
\begin{align}
(Q_{1}^{{\rm SLR},u})^{\dag} &
=  \widetilde O_{sduu}^{S,RL}\,,&
(Q_{2}^{{\rm SLR},u})^{\dag} &
= O_{sduu}^{S,RL}\,,\\
(Q_1^{{\rm VLR},d-s})^{\dag} & = -2O_{sddd}^{S,{LR}}+2O_{sdss}^{S,RL} \,,&
(Q_2^{{\rm VLR},d-s})^{\dag} & = O_{sddd}^{V,LR}-O_{sdss}^{V,LR}\,,
\end{align}
{\bf Class III}
\begin{align}
(Q_{1}^{{\rm SLL}, u})^{\dag} &
= \widetilde O_{sduu}^{S,RR}\, ,&
(Q_{2}^{{\rm SLL}, u})^{\dag} &
= O_{sduu}^{S,RR}\,,
\\
(Q_{3}^{{\rm SLL}, u})^{\dag} &
= -\widetilde O_{sduu}^{T,RR}\, ,&
(Q_{4}^{{\rm SLL}, u})^{\dag} &
= - O_{sduu}^{T,RR}\,,
\end{align}
{\bf Class IV}
\begin{align}
  (Q_{1}^{{\rm SLL}, d})^{\dag} &
  = -\frac{1}{2}O_{sddd}^{S,RR}-\frac{1}{8}O_{sddd}^{T,RR}\, ,&
  (Q_{2}^{{\rm SLL}, d})^{\dag} &
  = O_{sddd}^{S,RR}\, ,
    \\
      (Q_{1}^{{\rm SLL}, s})^{\dag} &
      = -\frac{1}{2}O_{sdss}^{S,RR}-\frac{1}{8}O_{sdss}^{T,RR}\,,&
        (Q_{2}^{{\rm SLL}, s})^{\dag} &
        = O_{sdss}^{S,RR}\,.
\end{align}
\subsection{From SD  to Flavio basis}\label{Flavio2}
Here we report the inverse transformation of the relations in the previous subsection.
\begin{align}
  O_{sduu}^{V,LL} & = \frac{1}{4}Q_1^{\dag}\, , &
\widetilde  O_{sduu}^{V,LL} & = \frac{1}{4}Q_2^{\dag}\, , \\
O_{sddd}^{V,LL} & = -\frac{1}{8}Q_1^{\dag}+\frac{1}{8}Q_3^{\dag}+\frac{1}{2}(Q_{1}^{{\rm VLL}, d-s})^{\dag}\, , &
O_{sdss}^{V,LL} & = -\frac{1}{8}Q_1^{\dag}+\frac{1}{8}Q_3^{\dag}-\frac{1}{2}(Q_{1}^{{\rm VLL}, d-s})^{\dag}\, ,
\end{align}

\begin{align}
  O_{sduu}^{V,LR} & = \frac{1}{12}Q_5^{\dag}+\frac{1}{6}Q_7^{\dag}\, , &
\widetilde  O_{sduu}^{V,LR} & = \frac{1}{12}Q_6^{\dag}+\frac{1}{6}Q_8^{\dag}\, , \\
O_{sddd}^{V,LR} & = \frac{1}{12}Q_5^{\dag}-\frac{1}{12}Q_7^{\dag}+\frac{1}{2}(Q_{2}^{{\rm VLR}, d-s})^{\dag}\, , &
O_{sdss}^{V,LR} & = \frac{1}{12}Q_5^{\dag}-\frac{1}{12}Q_7^{\dag}-\frac{1}{2}(Q_{2}^{{\rm VLR}, d-s})^{\dag}\, ,
\end{align}

\begin{align}
  O_{sduu}^{S,RL} & = (Q_{2}^{{\rm SLR}, u})^{\dag}\, , &
\widetilde  O_{sduu}^{S,RL} & = (Q_{1}^{{\rm SLR}, u})^{\dag}\, , \\
O_{sddd}^{S,{LR}} & = -\frac{1}{24}Q_6^{\dag}+\frac{1}{24}Q_8^{\dag}-\frac{1}{4}(Q_{1}^{{\rm VLR}, d-s})^{\dag}\, , &
O_{sdss}^{S,RL} & = -\frac{1}{24}Q_6^{\dag}+\frac{1}{24}Q_8^{\dag}+\frac{1}{4}(Q_{1}^{{\rm VLR}, d-s})^{\dag}\, ,
\end{align}

\begin{align}
  O_{sduu}^{S,RR} & = (Q_{2}^{{\rm SLL}, u})^{\dag}\, , &
\widetilde  O_{sduu}^{S,RR} & = (Q_{1}^{{\rm SLL}, u})^{\dag}\, , \\
O_{sddd}^{S,RR} & = (Q_{2}^{{\rm SLL}, d})^{\dag}\, , &
O_{sdss}^{S,RR} & = (Q_{2}^{{\rm SLL}, s})^{\dag}\, ,
\end{align}

\begin{align}
  O_{sduu}^{T,RR} & = -(Q_{4}^{{\rm SLL}, u})^{\dag}\, , &
\widetilde  O_{sduu}^{T,RR} & = -(Q_{3}^{{\rm SLL}, u})^{\dag}\, , \\
O_{sddd}^{T,RR} & = -8(Q_{1}^{{\rm SLL}, d})^{\dag}-4(Q_{2}^{{\rm SLL}, d})^{\dag}\, , &
O_{sdss}^{T,RR} & = -8(Q_{1}^{{\rm SLL}, s})^{\dag}-4(Q_{2}^{{\rm SLL}, s})^{\dag}\, .
\end{align}

\section{Meson evolution in the SD basis}\label{MEinSD}

Using the notation $\hat \Lambda = \Lambda/(4 \pi F)$ for short, we have the following evolution:
\\

{\bf Class I:}

\begin{equation}\label{eq:ALambdaEV}
  Q_1^{{\rm VLL},d-s}(\Lambda)=\left[1-4\hat\Lambda^2\right]Q_1^{{\rm VLL},d-s}(0)\,,
\end{equation}

 {\bf Class II}

\begin{align}
   Q_{1}^{{\rm SLR},u}(\Lambda) & = Q_{1}^{{\rm SLR},u}(0)+8\frac{M^2}{r^2}\hat\Lambda^2Q_{2}^{{\rm SLR},u}(0)\,,
\\
   Q_{2}^{{\rm SLR},u}(\Lambda) & =\left[1-\frac{4}{3}\hat\Lambda^2\right] Q_{2}^{{\rm SLR},u}(0)\,,
\\
    Q_1^{{\rm VLR},d-s}(\Lambda) & =\left[1-\frac{4}{3}\hat\Lambda^2\right]  Q_1^{{\rm VLR},d-s}(0)\,,
\\
    Q_2^{{\rm VLR},d-s}(\Lambda) & =  Q_2^{{\rm VLR},d-s}(0)+8\frac{M^2}{r^2}\hat\Lambda^2 Q_1^{{\rm VLR},d-s}(0)\,,
\end{align}

{\bf Class III}

\begin{align}
  Q_{1}^{{\rm SLL}, u} (\Lambda) & =\left[1+\frac{4}{3}\hat\Lambda^2\right] Q_{1}^{{\rm SLL}, u}(0)+4\hat\Lambda^2 Q_{2}^{{\rm SLL}, u}(0)\,,
\\
  Q_{2}^{{\rm SLL}, u} (\Lambda) & =\left[1+\frac{4}{3}\hat\Lambda^2\right] Q_{2}^{{\rm SLL}, u}(0)+2 \hat\Lambda^2 Q_{1}^{{\rm SLL}, u} (0)-\frac{1}{2}\hat\Lambda^2 Q_{3}^{{\rm SLL}, u} (0)\,,
\\
  Q_{3}^{{\rm SLL}, u} (\Lambda) & =\left[1+\frac{4}{3}\hat\Lambda^2\right] Q_{3}^{{\rm SLL}, u}(0)-16 \hat\Lambda^2 Q_{2}^{{\rm SLL}, u}(0)\,,
\\
  Q_{4}^{{\rm SLL}, u} (\Lambda) & =\left[-8  Q_{1}^{{\rm SLL}, u} (0)+2Q_{3}^{{\rm SLL}, u}(0)\right]\hat\Lambda^2\,,
\end{align}

{\bf Class IV}

\begin{align}
  Q_{1}^{{\rm SLL}, d} (\Lambda) & =\left[1-\frac{8}{3}\hat\Lambda^2\right] Q_{1}^{{\rm SLL}, d} (0)+\frac{4}{3}\hat\Lambda^2Q_{2}^{{\rm SLL}, d}(0)\,,
\\
  Q_{2}^{{\rm SLL}, d} (\Lambda) & =\left[1-\frac{8}{3}\hat\Lambda^2\right] Q_{2}^{{\rm SLL}, d} (0)\,,
\end{align}
with analogous equations for $ Q_{1,2}^{{\rm SLL}, s}$.

\section{Quark-gluon evolution in the SD basis}\label{SDADM}

The anomalous dimension matrices for all BSM operators are then given in the
SD basis as follows (in units of $\as/4\pi$) \cite{Buras:2000if,Aebischer:2017gaw}.
\newline
\newline
{\bf Class I}

\be
\hat\gamma^{(0)}(Q_{1}^{\rm VLL,d-s})= 4\,,
\ee
\newline
\newline
{\bf Class II}
\bea
\hat{\gamma}^{(0)}(Q_{1}^{\rm SLR,u}, Q_{2}^{\rm SLR,u}) &=& \left( \begin{array}{ccc}
\f{6}{N} && -6 \\[1mm]
0 && - 6N+ \f{6}{N}
\end{array} \right)=
\left( \begin{array}{ccc}
2 && -6 \\[1mm]
0 && - 16
\end{array} \right)\,,
\eea
\bea
\hat{\gamma}^{(0)}( Q_{1}^{\rm VLR,d-s}, Q_{2}^{\rm VLR,d-s}) &=& \left( \begin{array}{ccc}
-6 N +\f{6}{N} &~& 0 \\[1mm]
-6    && \f{6}{N}
\end{array} \right)= \left( \begin{array}{ccc}
-16  &~& 0 \\[1mm]
-6    && 2
\end{array} \right), \label{g045}
\eea
where $N$ denotes the number of colours with $N=3$.
\newpage
{\bf Class III}
\vspace{2mm}

In the basis $(Q_{1}^{\rm SLL,u},Q_{2}^{\rm SLL,u}, Q_{3}^{\rm SLL,u}, Q_{4}^{\rm SLL,u})$ we have
\bea
\hat{\gamma}^{(0)\rm SLL,u} &=& \left( \begin{array}{cccc}
\f{6}{N} & -6 & \f{N}{2}-\f{1}{N} & \f{1}{2} \\
0 & - 6N+\f{6}{N} & 1 & -\f{1}{N} \\
-\f{48}{N} + 24N & 24 & -\f{2}{N} - 4N & 6 \\
48  & -\f{48}{N} & 0 & 2N-\f{2}{N}
\end{array} \right)\\
&=&
\left( \begin{array}{cccc}
2 & -6 & 7/6 & {1}/{2} \\
0 & - 16 & 1 & -{1}/{3} \\
56 & 24 & -38/3 & 6 \\
48  & -16 & 0 & 16/3
\end{array} \right).
\label{ga0SLL}
\eea

{\bf Class IV}

\bea
\hat{\gamma}^{(0)}(Q_{1}^{\rm SLL,d},Q_{2}^{\rm SLL,d}) &=& \left( \begin{array}{ccc}
 2 N+4+\f{2}{N} &~&  4 N-4-\f{8}{N}  \\[1mm]
  4-\f{8}{N} &&  -{6}{N}+8+\f{2}{N} \end{array} \right)\\
 &=&
\left( \begin{array}{ccc}
 32/3 &~&  16/3  \\[1mm]
  4/3 &&  -28/3 \end{array} \right)\,,
\eea
with the same matrix for the operators $Q_{1,2}^{\rm SLL,s}$.

\section{Hadronic matrix elements in the SD basis}\label{sec:BBM}

In the large-$N$ limit the matrix elements of the two most important SM operators
are given as follows  \cite{Buras:1985yx,Bardeen:1986vp,Buras:1987wc}
\be\label{eq:Q60}
\langle Q_6(\mu) \rangle_0 =-\, r^2(\mu)  (F_K-F_\pi)\,,\qquad \langle Q_6(\mu) \rangle_2=0\,,
\ee
\be\label{eq:Q82}
\langle Q_8(\mu) \rangle_0 = \frac{1}{2}
 r^2(\mu) F_\pi\,, \qquad \langle Q_8(\mu) \rangle_2 =\frac{1}{2\sqrt{2}}
r^2(\mu) F_\pi \,.
\ee

The matrix elements of BSM operators are listed below.  We omit for brevity
the argument $\Lambda=0$.
\newline

{\bf Class I}

\begin{align}
\langle Q_{1}^{\rm VLL,d-s}\rangle_0 =&   +\f{F_\pi}{12}(m_K^2-m_\pi^2)\,,
&
\langle Q_{1}^{\rm VLL,d-s}\rangle_2 =&   -\sqrt{2}\,\f{F_\pi}{12}(m_K^2-m_\pi^2)\,.
\end{align}

\newpage

{\bf Class II}

\begin{align}
\langle Q_{1}^{\rm SLR,u}\rangle_0 &= {{-}}\f{F_\pi}{12}(m_K^2-m_\pi^2)\,,
&
\langle Q_{1}^{\rm SLR,u}\rangle_2 &= {{-}}\f{F_\pi}{12\sqrt{2}}(m_K^2-m^2_\pi)\,,
\\
\langle Q_{2}^{\rm SLR,u}\rangle_0 &= -\f{1}{12} r^2(\mu) F_\pi\,,
&
\langle Q_{2}^{\rm SLR,u}\rangle_2 &= +\f{1}{24\sqrt{2}} r^2(\mu) F_\pi
\,,
\\
\langle Q_{1}^{\rm VLR,d-s}\rangle_0 &=  - \f{1}{12} r^2(\mu) F_\pi\,,
&
\langle Q_{1}^{\rm VLR,d-s}\rangle_2 &=  - \f{1}{12\sqrt{2}} r^2(\mu) F_\pi\,,
\\
\langle Q_{2}^{\rm VLR,d-s}\rangle_0 &=   {-}\f{F_\pi}{12}(m_K^2-m_\pi^2)\,,
&
\langle Q_{2}^{\rm VLR,d-s}\rangle_2 &=  {+}\sqrt{2}\,\f{F_\pi}{12}(m_K^2-m_\pi^2)\,.
\end{align}

{\bf Class III}

\begin{align}
\langle Q_{1}^{\rm SLL,u}\rangle_0 &=  +\f{1}{48} r^2(\mu) F_\pi\,,
&
\langle Q_{1}^{\rm SLL,u}\rangle_2 &=  +\f{1}{48\sqrt{2}} r^2(\mu) F_\pi\,,
\\
\langle Q_{2}^{\rm SLL,u}\rangle_0 &=  - \f{1}{24} r^2(\mu) F_\pi\,,
&
\langle Q_{2}^{\rm SLL,u}\rangle_2 &=  - \f{1}{24 \sqrt{2}} r^2(\mu) F_\pi\,,
\\
\langle Q_{3}^{\rm SLL,u}\rangle_0 &=  - \f{1}{4} r^2(\mu) F_\pi\,,
&
\langle Q_{3}^{\rm SLL,u}\rangle_2 &=  - \f{1}{4\sqrt{2}} r^2(\mu) F_\pi \,,
\\
\langle Q_{4}^{\rm SLL,u}\rangle_0 &=  0\,,
&
\langle Q_{4}^{\rm SLL,u}\rangle_2 &=  0\,.
\end{align}

{\bf Class IV}

\begin{align}
\langle Q_{1}^{\rm SLL,d}\rangle_0 &=  +\f{1}{24} r^2(\mu) F_\pi\,,
&
\langle Q_{1}^{\rm SLL,d}\rangle_2 &=  -\f{1}{48\sqrt{2}} r^2(\mu) F_\pi\,,
\\
\langle Q_{2}^{\rm SLL,d}\rangle_0 &=  - \f{1}{12} r^2(\mu) F_\pi\,,
&
\langle Q_{2}^{\rm SLL,d}\rangle_2 &=  +\f{1}{24\sqrt{2}} r^2(\mu) F_\pi \,.
\end{align}

\renewcommand{\refname}{R\lowercase{eferences}}

\addcontentsline{toc}{section}{References}

\bibliographystyle{JHEP}
\bibliography{Bookallrefs}
\end{document}